\newif\iflandscape
\newif\ifportrait
\portraittrue
\newlength{\extralineskip}
\ifportrait
  \documentstyle[12pt]{article}
\fi
\iflandscape
  \documentstyle[twocolumn]{article}
\fi
\def\tr#1{{\rm tr}\kern-3pt\left[#1\right]}

\def\Tr{{\rm Tr}}
\makeatletter
\newdimen\normalarrayskip              
\newdimen\minarrayskip                 
\normalarrayskip\baselineskip
\minarrayskip\jot
\newif\ifold             \oldtrue            \def\new{\oldfalse}
\def\arraymode{\ifold\relax\else\displaystyle\fi} 
\def\eqnumphantom{\phantom{(\theequation)}}     
\def\@arrayskip{\ifold\baselineskip\z@\lineskip\z@
     \else
     \baselineskip\minarrayskip\lineskip2\minarrayskip\fi}
\def\@arrayclassz{\ifcase \@lastchclass \@acolampacol \or
\@ampacol \or \or \or \@addamp \or
   \@acolampacol \or \@firstampfalse \@acol \fi
\edef\@preamble{\@preamble
  \ifcase \@chnum
     \hfil$\relax\arraymode\@sharp$\hfil
     \or $\relax\arraymode\@sharp$\hfil
     \or \hfil$\relax\arraymode\@sharp$\fi}}
\def\@array[#1]#2{\setbox\@arstrutbox=\hbox{\vrule
     height\arraystretch \ht\strutbox
     depth\arraystretch \dp\strutbox
     width\z@}\@mkpream{#2}\edef\@preamble{\halign \noexpand\@halignto
\bgroup \tabskip\z@ \@arstrut \@preamble \tabskip\z@ \cr}%
\let\@startpbox\@@startpbox \let\@endpbox\@@endpbox
  \if #1t\vtop \else \if#1b\vbox \else \vcenter \fi\fi
  \bgroup \let\par\relax
  \let\@sharp##\let\protect\relax
  \@arrayskip\@preamble}
\def\eqnarray{\stepcounter{equation}%
              \let\@currentlabel=\theequation
              \global\@eqnswtrue
              \global\@eqcnt\z@
              \tabskip\@centering
              \let\\=\@eqncr
              $$%
 \halign to \displaywidth\bgroup
    \eqnumphantom\@eqnsel\hskip\@centering
    $\displaystyle \tabskip\z@ {##}$%
    &\global\@eqcnt\@ne \hskip 2\arraycolsep
         $\displaystyle\arraymode{##}$\hfil
    &\global\@eqcnt\tw@ \hskip 2\arraycolsep
         $\displaystyle\tabskip\z@{##}$\hfil
         \tabskip\@centering
    &{##}\tabskip\z@\cr}
\makeatother
\def\theequation{\thesubsection.\arabic{equation}}

\def\f{1\over }
\def\beq{\begin{equation}}
\def\eeq{\end{equation}}
\def\bea{\begin{eqnarray}}
\def\eea{\end{eqnarray}}
\def\be{\begin{eqnarray}}
\def\ee{\end{eqnarray}}
\def\nn{\nonumber}
\def\mm{matrix model }
\def\mms{matrix models }
\def\ba{\beq\new\begin{array}{c}}
\def\ea{\end{array}\eeq}
\def\stackreb#1#2{\mathrel{\mathop{#2}\limits_{#1}}}
\def\Bf#1{\mbox{\boldmath $#1$}}

\def\balpha{{\Bf\alpha}}
\def\bbeta{{\Bf\beta}}

\def\bmu{{\Bf\mu}}
\def\bphi{{\Bf\phi}}

\def\bn{{\Bf n}}

\def\bJ{{\Bf J}}
\def\be{{\Bf e}}

\def\bsalpha{{\Bf\alpha}}
\def\bsbeta{{\Bf\beta}}

\def\bsphi{{\Bf\phi}}

\def\bsn{{\Bf n}}

\def\W{{\rm W}}

\begin{document}

\begin{titlepage}
\setcounter{footnote}0
\begin{center}
\hfill FIAN/TD-24/93\\
\hfill UBC/S-95/93\\
\hfill hep-th/9312212\\
\begin{flushright}{September 1993}\end{flushright}
\vspace{0.1in}{\LARGE\bf $2d$ gravity and matrix models}\\
[.2in]
{\Large I. $2d$ gravity
}\footnote{Based on lectures given at the XXVII Winter LIYaF School,
February, 1993}
\\[.4in]
{\Large A.Mironov\footnote{E-mail address:
mironov@td.fian.free.net, mironov@grotte.teorfys.uu.se}}\\
\bigskip {\it Theory Department ,  P.N.Lebedev Physics
Institute , Leninsky prospect, 53, Moscow,~117924, Russia}
\end{center}
\bigskip
\bigskip

\centerline{\bf ABSTRACT}
\begin{quotation}
\noindent
Some approaches to $2d$ gravity developed for the last years are
reviewed. They are physical (Liouville) gravity, topological theories and
matrix models. A special attention is paid
to matrix models and their interrelations
with different approaches. Almost all technical details are omitted, but
examples are presented.
\end{quotation}

\end{titlepage}
\clearpage

\newpage
\tableofcontents
\newpage
\setcounter{page}2
\setcounter{footnote}0
\section{Introduction}
After Polyakov proposed in 1981 the way to deal with non-critical strings
\cite{Polyakov}, there were many attempts to get some concrete results.
However,
it
still remains one of the most important and challenging problems in modern
string theory. Indeed, instead of ordinary (string) matter theory Polyakov
considered
the theory of matter fields coupled to two-dimensional gravity. This is why the
investigation of such theories is of great importance.

Actually, the revival of the interest to these theories occured
after 1988, mainly due to
the
calculation of the critical indices proposed by some groups both in light
cone \cite{KPZ} and in conformal \cite{DDK} gauges. These groups have given
strong evidence in favor of the existence of self-consistent non-critical
strings out of "magic" interval of the central charges of the matter
fields of the theory  $1\le c\le 25$.

The next serious sucsess was achieved in the fall of 1989 when there were
invented some new "non-field" viewpoints to the theory of $2d$ gravity
which have given efficient tools to deal with the matter theories with the
cental charge $c\le 1$. Below I try to review this new development
as it is observed from nowdays.

Indeed, in section 2  I just sketchily describe the "field theory" approach
of \cite{KPZ,DDK}, i.e. Liouville field theory. But this approach, which I
refer to "physical gravity" is still poorly developed, so the next sections are
devoted to some
new frameworks like topological
theories of different types (section 3) and matrix models (sections 4 and
5).
Some concluding remarks can be found in section 6.
Let us emphasize that we intentedly present (in section 5) only sketchy
review of matrix models in external fields. This very important approach
deserves a separate paper. We hope to present it in the second part of the
review, which is to be devoted to matrix models in applications to wider set
of physical problems (like gauge theories).

In fact, the main part of the review is devoted to matrix model approach, the
first, as it is the main point of authors's interest and, the second, as just
within this approach the most striking, powerful and general results were
obtained.
This review is goaled to non-specialists, therefore, almost all tedious
technical details are omitted, but the text is provided by a number of examples
and simple calculations. To get more detailed discussion, one can use
references to original papers presented in all necessary places.
I would like to stress that throughout the paper I mainly follow the approach
developed by our group. Therefore, the references to different approaches
are poorly presented. They can be found in two very detailed recent reviews
\cite{GiMo,dFGZ-J}.

\section{Physical gravity}
\subsection{Liouville theory}
\setcounter{equation}{0}
\subsubsection{Classical Liouville theory}
In this section we will briefly consider the Liouville theory having in minds
that it describes, being properly interpreted, the theory of physical $2d$
gravity. To begin with, let us consider the theory with the canonical Liouville
action:

\beq\label{Laction}
S_L={\f 4\pi}\int d^2z\left[{\f 2}\partial\phi\bar\partial\phi-{\f \beta^2}
e^{\sqrt{2}\beta\phi}\right],
\eeq
where the field depends on variables $z,\bar z$, standard complex variables
in the Euclidean space-time and $\partial\equiv{\partial\over \partial z}$,
$\bar\partial\equiv{\partial\over\partial\bar z}$.
This theory has two main features - it possesses conformal invariance at
the classical level and it has no constant-valued ground state solution, which
results in disastrous infra-red singularities.

As for the first property, naively the theory is not conformal, since the
trace of the stress tensor is equal to $T_{z\bar z}={\f \beta^2}e^{\sqrt{2}
\beta\phi}$.
The chiral component of the stress tensor for the action (\ref{Laction}) is

\beq
T_{zz}=-{\f 2}(\partial\phi)^2
\eeq
and so for $T_{\bar z\bar z}$. However, it is not holomorphic, even on the
equations of motion, since

\beq
\bar\partial T_{zz}={\sqrt{2}\over\beta}e^{\sqrt{2}\beta\phi}\partial\phi.
\eeq
This naive stress tensor can be improved by adding the term which does not
effect to the conservation law:

\beq
T_{\mu\nu}\to
T_{\mu\nu}-(\partial^2\delta_{\mu\nu}-\partial_{\mu}\partial_{\nu}
)\Sigma,\ \ \ \ T=T_{zz}+\partial^2\Sigma.
\eeq
Now, taking $\Sigma={\f\sqrt{2}\beta}\phi$, we get

\beq\label{imptensor}
T=-{\f 2}(\partial\phi)^2+{\f \sqrt{2}\beta}\partial^2\phi
\eeq
and

\beq
\bar\partial T={\sqrt{2}\over \beta}e^{\sqrt{2}\beta\phi}\partial\phi +
{\f \sqrt{2}\beta}\partial^2\bar\partial\phi=0,
\eeq
with using the equations of motion.

In fact, $\Sigma$ adds to the action just full derivative, proportional to
$\partial\bar\partial\phi$, which might be essential only when considering
non-trivial boundary conditions like the presence of vacuum
charge \cite{DF,Dotsenko,MarMir}.
This term in action can be invariantly written as

\beq\label{cterm}
{\sqrt{2}\over\beta}R\phi,
\eeq
where $R=\partial\bar\partial\log \rho$, $\rho=g_{z\bar z}$ are the
two-dimensional curvature and metric respectively. The Poisson brackets
of the stress tensor (\ref{imptensor})
produce the Virasoro algebra\footnote{It is
evident from vanishing of trace of the improved stress tensor on the equations
of motion.} with conformal anomaly presented already at the classical level:

\beq\label{anomaly}
c_{cl}={6\over \beta^2}.
\eeq

\subsubsection{Quantum Liouville theory}
Now let us treat the quantum corrections to the Liouville theory. Indeed, the
theory is ill-defined due to infra-red singularities. Nevertheless, one can try
to calculate the quantities not sensitive to the infra-red divergences. Say,
one
can consider ultra-violet renormalizations of the action (\ref{Laction})
with the curvature term (\ref{cterm}).

Indeed, it is not difficult to take into account all ultra-violet divergent
diagramms, as in two dimensions these are only tadpoles. The calculation is
rather trivial, and results in multiplicative renormalization of the coupling
constant ${\f \beta^2}$ of the exponential in the action (\ref{Laction})
by the factor $\Lambda^{-2\beta^2}$. It is the same as
shift of the vacuum expectation value of field $\phi$ by the quantity
$\delta=-\sqrt{2}\beta\log \Lambda$. Let us note that this
shift depends on ultra-violet cut-off parameter $\Lambda$ which should be made
dimensionless by
some infra-red cut-off parameter. This parameter can be naturally chosen to be
the order of magnitude of the
metric $\rho$, as it is metric that cuts-off the theory at large
distances in curved space\footnote{It again can be turned to the matter of
boundary conditions.}. Thus, we can put $\phi \longrightarrow \phi -
\sqrt{2}\beta\log
\Lambda \rho$. It just results in the renormalization of the coefficient
in the curvature term (\ref{cterm}):

\beq\label{rencoef}
{\f 2}\partial\phi\bar\partial\phi \rightarrow - {\f 2}(\phi+\delta)
\partial\bar\partial(\phi+\delta)={\f 2}\partial\phi\bar\partial\phi-\phi
\partial\bar\partial\delta = {\f 2}\partial\phi\bar\partial\phi +
\sqrt{2}\beta R\phi.
\eeq
It results in the cental charge equal to

\beq\label{ccL}
c=1+6(\beta+{\f \beta})^2.
\eeq

Now let us justify (or illuminate) this result. Let us suggest that one should
still
preserve the conformal invariance at the quantum level. It means that the
exponential term in the action should be of the dimension (1,1), i.e. should
have
canonical dimension. From the other hand, the conformal theory with the action
(\ref{Laction}) and additional $\alpha R\phi$ term
has the central charge \cite{Dotsenko,MarMir} $c=1+3\alpha^2$, and the "naive"
dimension $\Delta_{\gamma}={\gamma^2\over 2}$
of the operator $:\exp\{\gamma\phi\}:$~\footnote{Throughout the review we
omit the sign of normal ordering where this can not be misleading.},
which it has in the gaussian model, changes
to $\Delta_{\gamma}={\f 2}\gamma(\alpha-\gamma)$.
As this dimension is to be unit, one gets the expression for the coefficient
$\alpha$:

\beq\label{alpha}
\alpha=\sqrt{2}(\beta+{\f \beta}),\ \beta={\alpha\over\sqrt{8}} \pm
\sqrt{{\alpha^2\over 8}-1},
\eeq
which coincides with that in the formula (\ref{rencoef}).
It leads to the same expression for the central charge (\ref{ccL}).

Certainly, this consideration was rather a motivation than the rigid result.
Indeed, the last reasoning is not too correct, as we treated only chiral
objects throughout the discussion, but one can separate chiral and
anti-chiral parts only on the equations of motion. This approach can be done
completely correct only in the case of $N=2$ Liouville theory \cite{MarMor}.

\subsubsection{Sin-Gordon viewpoint}
Now we will say some words on how to deal with the
Liouville theory in a more correct
way. Let us note that the exact computation of all ultra-violet contribution
would lead to the complete final answer in the
absence of infra-red divergences.
Indeed, there are no dimensional parameters, excluding UV cut-off parameter,
and, therefore, any non UV contributions to the effective action
are forbidden by the dimension argument.
Unfortunately, this is not the case in the
Liouville theory as it has terrible infra-red divergences, and, therefore,
essentially depends on the IR cut-off parameter. These IR divergences are due
to
the absence of finite-density constant classical solution of the theory, i.e.
due to the absence of classical vaccum in the theory. In principle, this
ill-defined behaviour should be expected from the very begining as the theory
is the conformal one, and, therefore, has no good particle content, finite
$S$-matrix and etc. It implies that, to deal with the Liouville theory, one
should consider this as the theory given in a fixed point
 of some larger theory. The natural
choices of this larger theory are the sin- or sinh-Gordon theories, depending
on whether $\beta$ is real or pure imaginary\footnote{The consideration above
was
equally correct for any complex $\beta$.}. It was investigated
in the paper \cite{Grisaru}, where the authors
worked with the sin-Gordon theory (i.e. $\beta\equiv i\tilde\beta$ is pure
imaginary) with the action\footnote{It is essential
for calculations to have {\it two} different constants $\gamma_+,\ \gamma_-$
in action, though one of them can be removed by a shift of the field. This
shift is often infinite in the cases under consideration.}:

\beq
S_{SG}={\f 4\pi}\int d^2z\left[\partial\phi\bar\partial\phi + \gamma_+
e^{i\sqrt{2}\tilde\beta\phi}+\gamma_-e^{-i\sqrt{2}\tilde
\beta\phi}+i\tilde\alpha
R\phi\right].\label{SGaction}
\eeq
For the sake of convenience, we also replaced $\alpha\to\tilde\alpha\equiv -i
\alpha$ in the action, i.e. $c=1-3\tilde\alpha^2$.
The authors of \cite{Grisaru}
 generalized field theory calculation above to this more complicated
case and have calculated $\beta$-functions perturbatively in $\gamma_{\pm}$
and $\tilde\beta^2-1$. They have observed that $\tilde\alpha/\tilde\beta$
is a RG
invariant and calulated $\beta$-functions. These were:

\ba\label{beta}
{d\lambda\over dt}\equiv \beta_{\tilde\beta}= 2
\tilde\beta^4\gamma_+\gamma_-,\\
{dq^2\over dt}\equiv \beta_{\tilde\alpha}
 = {\f 16}\tilde\alpha^2\tilde\beta^2\gamma_+\gamma_-,\\
{d\gamma_{\pm}\over dt}\equiv \beta_{\pm} = \gamma_{\pm}(\tilde\beta^2
 -
1 \mp {\f \sqrt{2}}\tilde\alpha\tilde\beta).
\ea
The theory appears to have both an infrared and an ultraviolet fixed points in
the region of approximation. There are two essentially different regimes which
were realized. One of them corresponds to RG flow from the free field theory
with a background charge $\tilde\alpha_+$
to the free field theory with a different
(and larger, in complete agreement with Zamolodchikov's $c$-theorem \cite{Zam})
background charge $\tilde\alpha_-$.
In the other regime, one flows from an UV fixed point
with $\gamma_+=\gamma_-=0$ to an IR fixed point, where $\gamma_-=0$, but
$\gamma_+\ne 0$, i.e. one
of the exponentials (marginal operator) survives in the
IR limit, leading to the Liouville action.

{}From the equation (\ref{beta}) one can determine the values of parameters
corresponding to this fixed point:

\beq
\tilde\beta={\tilde\alpha\over\sqrt{8}}\pm\sqrt{{\tilde\alpha^2\over 8}
+1}.
\eeq
This is consistent with the connection (\ref{alpha}), thus
justifying our results.

The manifest calculation of \cite{Grisaru} also reproduced the
values of dimensions which were
obtained above, and the generalized central charge,
which on the RG trajectory is given by some monotonic
decreasing function (in accordance with Zamolodchikov's $c$-theorem \cite{Zam})
at the fixed point describing Liouville theory reaches the value (\ref{ccL}).

\subsection{Liouville gravity}
\setcounter{equation}{0}
\subsubsection{Polyakov-KPZ-DDK approach}
In the previous section we considered Liouville theory with no reference
to the main purpose of the review, that is, the decription of $2d$ physical
gravity. In the section we are going to discuss $2d$ physical gravity itself.

One should start from the original covariant Polyakov action for the string
in flat $D$-dimensional euclidean target space:

\beq\label{straction}
S={\f 4\pi}\int d^2\xi \sqrt{g}g^{\mu\nu}\partial_{\mu}X^i\partial_{\nu}X^i,
\ \ i=1,...,D.
\eeq
The main purpose is to calculate the correlation functions through the
following path integral:

\beq\label{mainint}
<V_1V_2...V_n> \equiv \int {\cal D}X{\cal D}g^{\mu\nu}e^{-S}
\eeq
with the action (\ref{straction}).
We will proceed quantizing this system in conformal gauge, though there is
an equivalent light-cone gauge approach \cite{KPZ}. The conformal gauge
approach was proposed in the original paper by Polyakov \cite{Polyakov}, and
was developed in \cite{DDK}, where the central charge and weights of operators
were obtained.

Thus, we fix conformal gauge $g_{\mu\nu}=e^{\phi}\hat g_{\mu\nu}$, where
$\hat g_{\mu\nu}$ is a background metric. The gauge fixing procedure also
introduces reparameterization ghosts $b$, $c$, which are almost inessential
in our consideration. Therefore, the complete action consists of three pieces:
the ghost action, the
Liouville action (\ref{Laction}) with the proper curvature term
and the matter field action. As the complete action should be conformally
invariant and should not depend on the background metric $g_{\mu\nu}$, the
complete central charge must vanish\footnote{This means that the theory
ought to be topological one, see the next section.}.

In fact, the ghost stress tensor contributes -26 into the full central charge,
and the matter fields do $D$. Certainly, one can consider, instead of
(\ref{straction}), an arbitrary conformal matter theory, with an arbitrary
central charge $c_m$ instead $D$. We will do so to keep the most general
expressions admissing non-integer values of the central charge.

Now we can calculate the restriction due to zero full central charge, using
also the manifest expression for the Liouville central charge (\ref{ccL}):

\beq
c_m+c_{gh}+c_L=0\ \to \ c_m+(-26)+(1+3\alpha^2)=0,
\eeq
i.e.

\beq\label{alphaL}
\alpha=\sqrt{{25-c_m\over 3}},
\eeq
and $\alpha$ was determined in the previous subsection.

It is evident from the expression (\ref{alpha}) that the condition $
\alpha^2<8$ leads to the complex values of $\beta$ and the Liouville central
charge $c_L$ (\ref{ccL}), i.e. to non-unitary theory. It reproduces notorious
limitation on the central charge of matter (i.e. on the dimension of the
target space):

\beq
c_m\le 1.
\eeq
In fact, complex $\beta$ just means that the chosen vacuum of the theory
is unstable, and this vacuum is to pass to a different ground state. It can be
understood also from the tachyon mass square which is equal
to ${1-c_m\over 12}$. This means that the tachyon is absent and the vacuum
state is stable only provided by the same condition $c_m\le 1$.

There is another limit when $c_L$ is also real. It corresponds to
$\alpha^2<0$, i.e. $c\ge 25$. This implies that $\beta$ is pure imaginary,
or, what
is the same after redefiniton $\phi\to i\phi$, the wrong sign of the
kinetic Liouville term. It seems to spoil the unitarity of the theory too.

But there is a special case of $c_m=25$, when one just reproduces the critical
string with 25 space coordinates (matter fields) and one time coordinate
with reversed sign in the kinetic term (Liouville field). Therefore,
in such a way, we
recover the mechanism to get Minkowski space from Euclidean theory, which is
applicable for the critical string! Simultaneously it means that the
"physical" dimension
of non-critical string is $c_m+1$.

\subsubsection{Dressing of operators}
Now we are going to discuss operator content of $2d$ gravity. As the first
step,
let us note that the functional integral in gravity theory should be
independent
both of background metric $\hat g_{\mu\nu}$ and of physical metric $g_{\mu\nu}=
\hat g_{\mu\nu}e^{\sqrt{2}\beta\phi}$.\footnote{This renormalization of the
exponential factor - or the Liouville field - is a direct consequence of our
requirement for this exponential to have the dimension (1,1). The same
phenomenon can be described as well by saying that the scalar Liouville field
changes the trasformation law and is not scalar after taking into account the
conformal anomaly (see more detailed discussion in \cite{MarMor}).}

The first property implies zero full central charge of the theory and
was already used to connect Liouville exponential with the matter central
charge
$c_m$. The second requirement was used in the previous subsection to
calculate central charge of the Liouville theory (\ref{ccL}).

Now we will consider matter field in the same context. These fields
should be also
renormalized (dressed) by the Lioville field due to the interaction
with gravity. To calculate this dressing, we require for correlation functions
of the operators to be independent of metric,
instead of doing very tedious perturbative calculations.

Thus, let us consider a matter operator $V$ which has a dimension $\Delta$.
It is used to suggest that, as a result of gravity interaction, it is dressed
by
exponential of the Liouville field:

\beq\label{operator}
V_{dr}=e^{\gamma\phi}V.
\eeq

Actually, we would have to consider more general operators which include
Liouville secondaries (and maybe even ghost contributions).
Fortunately, it is not necessary to do in the $2d$ gravity case. This is due to
huge invariance of the theory.
 The drastic reduction of the space of vertex operators is a
general phenomenon occurring whenever a string model is build from CFT. Let us
remind first the situation with the critical string. In the simplest treatment,
one neglects the
Liouville field at all, but this is not the only thing one does.
There are three additional requirements to physical vertex operators in
critical string models:

a) They do not contain ghost fields (in certain and natural
picture)\footnote{Except for one delicate point, concerning dilaton operator.}.

b) They are Virasoro primaries (with respect to ``full" Virasoro). (All the
descendants, present in CFT are ``gauged out". The reason is that coupling to
$2d$ gravity implies gauging the Virasoro algebra and thus all the descendants
are eliminated as gauge-non-invariant operators. They really decouple from
the correlators\footnote{It is just no-ghost theorem.}).

c) They are integrals (in the picture consistent with  a)) of operators of
conformal dimension one\footnote{Let
us note that other (equivalent) pictures may be obtained by
multiplication of dimension one integrands (constrained by a) and b)) with
ghost field instead of integration.}.

Naive generalization of these three principles to the case of non-critical
string has been suggested in \cite{DDK},
with the only modification made at the
point b), where one is supposed to take primaries of the full ({\it matter} +
{\it Liouville}) Virasoro algebra, in the form of ({\it matter
primaries})$\times$({\it Liouville  primaries}) and
Liouville primaries being pure
exponentials. It is, however, not quite true that such a simple generalization
is valid. A problem arises at the very beginning -- at point a). In order to
{\it derive} honestly this requirement one should prove that in every BRST
cohomology class (i.e. for every physical state)
there is a ghost free representative. According to
\cite{LianZu,BMP} this
is not true in non-critical case: there are two additional representatives of
the cohomology classes, which unavoidably (and non-trivially) contain ghost
fields. In the particular case of $c=1$ model, the first ones were interpreted
in [2] as a ground ring, another one (the ``ghost number two") still has no
nice interpretation. It seems, however, that the subsector, described by
principles a),b),c) is closed by itself under OPE (modulo descendants, $i.e.$
fields vanishing in the correlation functions). Another delicate point is that
the requirement c), $i.e.$ that the full dimension is one, permits two
different choices of Liouville primaries associated with a given matter
operator. In what follows we consider a subsector with one specific choice of
these two. This subsector also seems to be closed under OPE in the above sense.

Thus, we restrict ourselves to this subsector and consider only the operators
of the form (\ref{operator}),
more general discussion can be found in \cite{MMMO}.
Let us calculate the value of the exponent $\gamma$ in (\ref{operator})
requiring (1,1) dimension of the dressed operator (it was already explained in
different words that it just expresses the possibility to integrate it
over $2d$ surface with no breaking conformal invariance). Then we obtain:

\beq
\Delta+{\f 2}\gamma[\alpha-\gamma]=1,
\eeq
i.e.

\beq\label{gamma}
\gamma={\f 2}(\alpha\pm\sqrt{\alpha^2-8+8\Delta}).
\eeq
Now let us say some words on the choice of sign in this expression (and that
in the Liouville exponent, see (\ref{alpha})). Hereafter, our prescription
will be to choose "-", because of arguments proposed in \cite{Sei}. There was
argued that this choice selects out the local operators corresponding to the
states with positive energy. We will comment more this point a bit later.

In the next subsection we use this expression to determine scaling
properties of dressed operators.

\subsubsection{String susceptibility}
As correlation functions of physical operators in $2d$ gravity do not depend
on metric, as well as on the positions of operators (since the integration
over locations of all operators is to be presented), all correlation
functions are just numbers. We will return to the question in the next section
and now let us only note that we still need some characteristic like
dimension, which differs among operators (and correlation functions). The
simplest way to introduce such a characteristic is to fix the area of the
surface $A$ and investigate $A$-dependence of correlation functions.
It is clear that this characteristic is of great importance, as it is that
which governs the grow (or shrinking) of the surfaces and controls their
stability.
It is the general feature of all string theories that their partition functions
exponentially decrease with area. The index which really differs among string
theories and governs the behaviour of their ground states is pre-exponential
factor.

We will now calulate this factor. Let us consider the path integral for
$2d$ gravity (=non-critical string) theory with fixed area $A$:

\beq\label{Lpartf}
Z(A)\equiv \int{\cal D}\phi{\cal D}(\hbox{\rm matter}){\cal D}(
\hbox{\rm ghosts})
e^{-S_L}\delta\left(\int d^2\xi\sqrt{\hat g}e^{\sqrt{2}\beta\phi}-A\right),
\eeq
where we included matter and ghost action terms into the measures and took into
account that the area of the surface is nothing but the integral of
the physical metric $g_{\mu\nu}=\hat g_{\mu\nu} e^{\sqrt{2}\beta\phi}$ over
the surface.

Though the integral (\ref{Lpartf}) is very complicated, its $A$-dependence can
be easily determined. Indeed, let us do the substitution $\phi\to\phi+
{\f \sqrt{2}\beta}\log A$. Then, the exponential in the Liouville action turns
into the exponential term $e^{-\hbox{\rm const}\times
 A}$ in the functional integral
(due to the $\delta$-function in (\ref{Lpartf})). Therefore, the only
contribution of the action to the pre-factor is from the curvature term:

\beq
{\f 4\pi}\int d^2\xi\sqrt{\hat g}\alpha R\phi \to
{\f 4\pi}\int d^2\xi\sqrt{\hat g}\alpha R\phi + {\alpha
\over\sqrt{2}\beta}
\log A \left({\f 4\pi}\int d^2\xi\sqrt{\hat g}R\right).
\eeq
Measure in the path integral (\ref{Lpartf}) is evidently invariant with respect
to constant shifts. Then, using the identity $\delta(Ax)=A^{-1}\delta(x)$ and
the Gauss-Bonnet theorem  ${\f 4\pi}\int d^2\xi\sqrt{\hat g}R = 1-g$, where
g is the genus of the surface, one finally obtains\footnote{We do not
concretize the constant in the exponential because of its non-universal
nature, in contrast to the index $\gamma_{str}$.}:

\beq
Z(A)\sim A^{-{\alpha\over\sqrt{2}\beta}(1-g)-1}e^{-\hbox{\rm const}\times
 A} \equiv
A^{(\gamma_{str}-2)(1-g)-1}e^{-\hbox{\rm const}\times A},
\eeq
i.e.

\beq\label{Gstr}
\gamma_{str}= 2-{\alpha\over\sqrt{2}\beta}={\f 12}\left(c_m-1-\sqrt{(c_m
-25)(c_m-1)}\right).
\eeq
This index $\gamma_{str}$ is called string susceptibility\footnote{Let us
remark that $\gamma_{str}$ has a complex value when $1<c_m<25$, pointing
once again to "the magic interval" of values of $c_m$.}.

\subsubsection{Weights of operators}
In the previous subsection we have calculated the index describing the
dependence of the partition function on area $A$. Analogous dependence of
correlation functions can be also investigated, and defines the index
which is a proper characteristic
of operator in $2d$ gravity.
More concretely, the index, which is called the weight of operator, can be
determined from the normalized one-point correlation function:

\ba
<V>_A\equiv{\f Z(A)}\int{\cal D}\phi{\cal D}(\hbox{\rm matter}){\cal D}
(\hbox{\rm ghosts})\times\\ \times
e^{-S_L}\delta\left(\int d^2\xi\sqrt{\hat g}e^{\sqrt{2}\beta\phi}-A\right)
\int d^2\xi\sqrt{\hat g}V_{dr}\sim A^{1-h}.
\ea
Physically, this index controls the behaviour (growth) of string surface
when the operator is inserted. Therefore, the standard notions of relevant,
marginal or irrelevant operators still make sense. In particular, relevant
operators correspond to $h<1$ and dominate in the IR (large $A$) limit.

To calculate $h$, we do the same steps as in course of calculation of
the string susceptibility and use the manifest expression for the
dressing exponent (\ref{gamma}):

\beq
<V>_A\sim{A^{-{\alpha(1-g)\over\sqrt{2}\beta}-1+{\gamma\over\sqrt{2}\beta}}\over
A^{-{\alpha(1-g)\over\sqrt{2}\beta}-1}}=A^{{\gamma\over\sqrt{2}\beta}},
\eeq
i.e.

\beq\label{h}
h=1-{\gamma\over\sqrt{2}\beta}={\sqrt{1-c_m+24\Delta_V}-\sqrt{1-c_m}\over
\sqrt{25-c_m}-\sqrt{1-c_m}}.
\eeq
Thus, the classification of operators into relevant, marginal and irrelevant
ones does not change for dressed operators. Indeed, the conditions $\Delta=1$,
$\Delta>1$ and $\Delta<1$ immediately imply the same conditions for the
dressed weights $h$ in (\ref{h}).

Let us note that the expression (\ref{h}) is real and positive for $c_m\le 1$
provided the choice of sign is "-" in the expressions (\ref{gamma}) and
(\ref{alpha}). This is one more argument in favor of that choice.

Now, to have an explicit example, let us write down the values of
$\gamma_{str}$
and $h$ for the operators in minimal models \cite{BPZ,Dotsenko,MarMir}. The
central charge of $(p,q)$-minimal model is equal

\beq
c=1-{6(p-q)^2\over  pq},
\eeq
i.e.

\beq
\gamma_{str}=-{2|p-q|\over p+q-|p-q|}.
\eeq
The (Kac) spectrum of operator dimensions is given by

\beq
\Delta_{m,n}={(pm-qn)^2-(p-q)^2\over 4pq},\ \ 1\le m\le q-1,\ \ 1\le n\le p-1,
\eeq
i.e.

\beq
h_{m,n}={|pm-qn|-|p-q|\over p+q-|p-q|}.
\eeq
This expression should be compared with the indices calculated in matrix models
(the below).

\section{Topological theories}
\def\theequation{\thesection.\arabic{equation}}
\setcounter{equation}{0}
We could conclude from the discussion of the previous section that
correlation functions in the theory of $2d$ gravity have a very simple
structure. Indeed, if we do not fix the area of the surface, then all
correlation functions are just numbers, provided by vanishing of the
full central
charge of the theory. This means that it is a topological theory.

Indeed, there no variables which correlation functions could depend on,
since integration over metrics is done. Certainly, it implies that the theory
is in unbroken gravity phase.

Let us look at the same argument from a different viewpoint. The routine way
to do (generally, non-critical) string integral was usually to fix the
conformal factor and then to remove two remaining components of metric tensor
by making use of the general covariance of the string action (i.e. to
integrate out the diffeomorphism group). After doing this one remains with the
integral over moduli of the conformal (=complex) structures and with the
integral over conformal (Liouville) factor (in the case of non-critical
strings). It is approximately the procedure which was implied in the previous
section.

Now let us change the order of the integrations. Namely, one can integrate
over metrics from the very beginning. Then, we have an integral (correlation
function) which can depend only on the locations of the points (as any
metric dependence is already integrated out) and still possesses the general
covariance (which is again correct only in unbroken phase). It can be
evidently nothing but a number.

Thus, we are led to consider the theory of $2d$ gravity as a topological
theory. Certainly, there is plenty of different topological theories, but
it is naturally to consider only $2d$ theories. Indeed, there are only
two sucsessful attempts to formulate a topological theory appropriate for
the description of $2d$ gravity.

The first attempt was due to Witten \cite{Wittop}, who proposed the theory
which preserves the memory of the moduli space treating as correlation
functions some integrals over moduli space. More concretely, let us
consider the holomorphic linear bundle ${\cal F}_i$ with the fiber being the
cotangent space (complex one dimensional vector space) to the Riemann
surface $\Sigma$ at the marked point $x_i$. Then we can take as correlation
functions the intersection numbers of these bundles. It means that we should
integrate the first Chern classes of the exterior products of these bundles
over (Deligne-Mumford) compactification ${\cal M}_{g,n}$ of the moduli
space of the curve $\Sigma$ of the genus $g$ (see also \cite{Kon}):

\beq\label{topcor}
<\sigma_{d_1}...\sigma_{d_n}>\equiv\int_{{\cal M}_{g,n}}\prod_i c_1({\cal
F}_i)^{\wedge d_i}.
\eeq
These correlation functions are indeed numbers (positive rational)
which vanish unless $6g-6+2n=2\sum d_i$. They are related by a number of
recursion relations which are equivalent to the Virasoro constraints of
the Hermitean one-matrix model in the continuum limit \cite{DVV}. This allows
one to identify the two theories.
Unfortunately, it is not so immediate to generalize the construction to
multi-matrix model case, and the work in this direction is not finished yet
(see, however, \cite{DW,WitN}).

Another sucsessful construction of the appropriate topological theory was
based on the follwoing general idea. Let us consider the theory whose
stress tensor (or action) is anticommutator of something with a BRST
operator $Q$: $T=\{\ast,Q\}$, $Q^2=0$. Then, this stress tensor describes a
topological theory on the subspace of the physical states (i.e. those
defined by the condition $Q|phys>=0$):

\beq\label{stress}
T|phys>=0.
\eeq
In fact, since coordinate variations are generated by the stress tensor, any
variation of a correlator is propostional to the the same
correlator with insertion of $T$. Therefore, correlator of the physical
fields, due to the condition (\ref{stress}), does not depend on the
locations of the points and is a number.

A concrete realization of this general idea has been proposed in \cite{Li} (see
also \cite{LGMT}), where twisted $N=2$ superconformal theory was
investigated. It was formulated in abstract terms of OPE, with no any
Lagrangians. The Lagrangian formulation, namely, $N=2$ supersymmetric
Landau-Ginzburg theory has been proposed in \cite{LGMTV}.

The advantage of the theories of described type is that they can be
investigated in some details, and they are applicable to the description
of plenty of matrix models, in contrast to the theories of Witten's
type. Because of the lack of space we can not develop here all these ideas, and
refer
to the reviews and papers containing all necessary references
\cite{Dij,Los,LosPol}.

\section{Discrete matrix models - the first step}
\def\theequation{\thesubsection.\arabic{equation}}
\subsection{A role of matrix models}
\setcounter{equation}{0}
Now let us say that the resultes mentioned in the section 2 are the only ones
obtained in the framework of physical gravity. But in the Fall of 1989
a drastical progress in $2d$ gravity was achieved in some other, rather
uncommon for the field theorists, frameworks. In particular, an old idea
proposed in \cite{Dav} to replace the path integral
of $2d$ gravity over metrics (\ref{mainint})
 by the sum over all possible triangulations of surfaces (i.e.
reformulate it in target space terms) have led, after all, to what is called
nowdays matrix models. Really it consists of three different components -
double
scaling limit \cite{BK,DS,GM}, constraint algebras satisfied by the partition
functions of matrix models \cite{DVV,FKN} and integrability properties
\cite{Dou}.

Up to now, the approach relies only on the original "physical" arguments of
triangulation\footnote{which are so approximate that even can not give any
physical interpretation of multicritical points} with sucsessive taking the
(double scaling) continuum limit and on the comparison of the very poor
predictions obtained from the physical gravity with those within matrix model
treatment. These predictions include weights of operators in the
theory and string susceptebility, as well as the
calculations of 3-point functions (and some $n$-point functions in very simple
channels). In fact, it means that the comparison is at the level of the spectra
of the theories. It implies, more or less, that the matter of quantum measure
is still out of the scope of the present investigations in physical gravity. In
fact, though there are some distinguished measures in the gravity path integral
(induced by the free field representation, or Polyakov's one, or ...), it is
still necessary to fix somehow these as well as the normalization factor
depending on the genus of the (Riemann surface) string world sheet (i.e. on the
order of
the perturbation theory).

At this point, one can reverse the line of reasoning and consider matrix models
as
{\it a definition} of these entries. To be a good definition, such a theory
should be defined naturally over all genera simultaneously. Indeed, I will try
to
demonstrate below that matrix models have to do with the infinite Grassmannian,
which, in its turn, naturally describes the Universal Moduli Space (UMS, the
space of all Riemann surfaces of all genera). Therefore, matrix models do
really
give the example of the theory with required properties.

Indeed, we will see that the partition function of a matrix model is a
$\tau$-function and corresponds to a concrete point of the Grassmannian
(corresponding to a Riemann surface of infinite genus) which implies that {\it
after}
doing the path integral (\ref{mainint}) over UMS, one reproduces the answer
corresponding to a point of the same UMS (see also \cite{Knizh,MorUFN,UMS}).
It resembles the phenomenon realized
some years ago \cite{BIKS} in $2d$ quantum integrable systems, when the exact
quantum correlators satisfy classical integrable equations\footnote{The analogy
is even closer as in string theory one should integrate some objects having
$\tau$-function sense \cite{Saito} giving rise to matrix model partition
function which is a $\tau$-function again, and in \cite{BIKS} the authors
obtained the classical integrable equations for a functional of correlation
functions which was also a $\tau$-function.}.

Thus, it is expected that matrix models give a correct description of the
non-critical strings with $c<1$, therefore, better understanding of their
properties is necessary. I will try to describe briefly these properties in the
next two sections. Indeed, now we can do even more - we establish the
connection of matrix models with some other recent approaches to $2d$ gravity
which can be also solved up to the very end (in contrast to physical gravity).
It is possible to reach the proof of complete identity between all these
theories. This should not be very surprising as all these theories calculate,
in essence, the same cohomologies and reflect
different general properties of the same final answer.

Now we just start from the matrix integral, the standard arguments in favor of
it can be found in plenty of reviews \cite{GiMo,dFGZ-J,Mor,Mar}.
Let us note that all
the specific properties of matrix models will be demonstrated in this section
for the example of Hermitian one matrix model.  More complicated examples (in
this and the next sections) will be labelled by asterisque and can be omitted
without disturbing the main line of reasoning.
I omit many subtle points and
all evident (but maybe technically interesting) generalizations of the
considered examples, refering to the original papers. Many of these
generalizations can be also
found in two recent reviews \cite{Mor,Mar}.

\subsection{Integrability in matrix models}
\setcounter{equation}{0}
 \subsubsection{Toda chain equation}
 Let us consider the integral over Hermitian $n\times n$ matrix with
 arbitrary potential:

 \beq\label{1MM}
 {\cal Z}_n(t_k) \equiv \{Vol_{U(n)}n!\}^{-1}\int [dH] \exp\{-{\rm Tr}V(H)\}, \
 \ V(H) \equiv - \sum_{i=0}
 t_kH^k,
 \eeq
 where $[dH]$ is the proper Haar measure.
 The standard procedure of doing this integral is, indeed, to fix some simple
 potential (polynomial of a finite degree), to write down the equations
 satisfied by the partition function (usually there are few equations as the
 integral (\ref{1MM}) depends on some coefficients in the polynomial potential)
 and to take the double scaling limit in these equations. Instead, we consider
 the partition function (\ref{1MM}) as the function of {\it all} coefficients
 $t_k$'s and demonstarte that all the crucial properties of matrix models
 (integrability and constraint algebra) can be exposed {\it before} taking the
 continuum limit. Moreover, in a sense, these properties are easier observed in
 the case.

 Thus, to begin with, let us demonstrate that the partition function
(\ref{1MM})
 is nothing but the $\tau$-function of the Toda chain integrable hierarchy
 \cite{GMMMO,KMMOZ}. We can rewrite (\ref{1MM}) as the integral only over
 eigenvalues with angular variables integrated over:

 \beq\label{1MME}
 {\cal Z}_n(t_k) = (n!)^{-1}\int   \prod  _i dh_i \Delta ^2(h) \exp \{
 -\sum _{i,k}t_{_k}h^{_k}_i\}.
 \eeq
 Now one can apply the standard machinery of the orthogonal polynomials
 \cite{BIZ}. To do
 this, let us define the polynomials by the orthogonality condition

 \beq\label{OP}
 \langle i|j \rangle \equiv \langle P_i,P_j \rangle = \int
 P_i(h)P_j(h)e^{-V(h)}dh =
 \delta _{ij}e^{\varphi _i(t)},
 \eeq
 where $e^{\varphi _i(t)}$ are the norms to be determined, the polynomial
 normalization being fixed by the unit coefficient in the leading term:

 \beq\label{OPN}
 P_i(h) = \sum _{j\leq i}a_{ij}h^j , \ \
 a_{ii} = 1.
 \eeq
 Using these polynomials, one can rewrite (\ref{1MME}) as

 \beq\label{prod}
 {\cal Z}_n =  (n!)^{-1}\int   \prod  _i dh_i \det P_{k-1}(h_j) \det
 P_{l-1}(h_m) \exp \{
 -\sum _{i,k}t_{_k}h^{_k}_i\} = \prod ^{n-1}_{i=0}e^{\varphi _i(t)}.
 \eeq
 Now we are going to prove that this partition function is nothing but
 a $\tau$-function of Toda chain hierarchy. In fact, we derive only the first
 equation, refering to \cite{GMMMO} for the general proof and any details.

 To begin with, let us consider the scalar product $<n|h|m>$ for different $m$.
 It is evidently zero whenever $m-n\ne 0,\pm 1$ (e.g., $<n|h|n-2>=<n|n-1> +
 \sum_{k>1} c_k <n|n-k> =0$ by the conditions (\ref{OP}) and (\ref{OPN})).
 It means that

 \beq\label{RecRel}
 hP_n(h)=P_{n+1}(h) - p_nP_n(h) + R_nP_{n-1}(h),
 \eeq
 where the coefficients $p_n$ and $R_n$ should be determined. The latter can be
 trivially calculated using (\ref{RecRel}) twice: $\ <n-1|h|n>=<hP_{n-1},P_n> =
 <n|n>+\sum_{k>0} <n|n-k>= <n|n> = <P_{n-1},hP_n> = R_n<n-1|n-1>$, i.e.

 \beq\label{Rn}
 R_n(t) = e^{\varphi _n - \varphi _{n-1}}.
 \eeq
 Indeed, the recurrent relation (\ref{RecRel}) already hints at the Toda chain,
 as, introducing difference ${\rm {\bf L}}$-operator with the matrix elements

 \beq\label{Loper}
 ({\rm {\bf L}})_{mn}=\delta_{m,n-1} - p_n\delta_{m,n} + R_{n+1}\delta_{m,n+1}
 \eeq
 with eigenfunction (Baker-Akhiezer function) $\Psi_n(h) = \exp\{-{1\over
 2}V(h)\} P_N(h)$, one can trivially see in (\ref{RecRel}) the action of Lax
 operator of the Toda chain \cite{GMMMO}, where $h$ is, as usual, the spectral
 parameter:

 \beq\label{LaxRep}
 {\rm {\bf L}}\Psi_n(h) = h \Psi_n(h).
 \eeq
 Now to get the equations of Toda chain it is sufficient to obtain the correct
 action of the operator of the derivative with respect to the first time
 \cite{GMMMO}. But more illuminating way is to derive the Toda chain equations
 \cite{GMMMO}. As an example, we demonstrate here the derivation only of the
 first equation in the hierarchy. To do this, let us differentiate $<n|n>$ with
 respect to the first time using (\ref{OP}), (\ref{OPN}) and (\ref{RecRel}):

 \beq\label{p_n}
 {\dot \varphi}_n  e^{\varphi _n } = {\partial\over \partial t_1} <n|n> = 2
 \left< {\partial P_n\over  \partial t_1}, P_n\right> - <n|h|n> = -p_n
 e^{\varphi _n},
 $$
 $$
 {\rm i.e.}\ \ p_n = - {\dot \varphi}_n,
 \eeq
 where dot means the derivative with respect to the first time. Now we
 differentiate the same quantity once more and, using (\ref{OP}), (\ref{OPN}),
 (\ref{RecRel}), (\ref{Rn}) and (\ref{p_n}) and integrating by parts, we obtain
 finally the well-known Toda chain equation:

 \beq
 \new
 \begin{array}{c}
 {\ddot \varphi}_n  e^{\varphi _n }+ ({\dot \varphi}_n)^2 e^{\varphi _n }
 ={\partial ^2 \over \partial t_1^2} <n|n> =
 <n|h^2|n> - 2 \left<{\partial P_n\over \partial t_1},hP_n\right> =\\
 = \left(p_n^2 +R_{n+1} + R_n\right) e^{\varphi _n } - 2R_n e^{\varphi _n },
 \ea
  i.e.

 \beq
  {\ddot \varphi}_n = R_{n+1} - R_n =  e^{\varphi
 _{n+1}-\varphi _n} - e^{\varphi _n - \varphi _{n-1}}.\label{TodaEq}
 \eeq
 This equation can be rewritten in the Hirota bilinear form leading to the
 notion of $\tau$-function, key object in the theory of integrable hierarchies.
 In fact, let us introduce this by the formula

 \beq\label{tauTC}
  e^{\varphi_n(t)} \equiv {\tau_{n+1}(t)\over \tau_n(t)}.
 \eeq
 Then, one can rewrite (\ref{TodaEq}) in the Hirota bilinear form as

 \beq\label{Hirota}
 \tau _n(t) {\partial ^2\over \partial t^2_1}\tau _n(t) -
 \left( {\partial \tau _n(t)\over \partial t_1}\right) ^2 =
 \tau _{n+1}(t)\tau _{n-1}(t).
 \eeq
 Indeed, the main sense of the notion of $\tau$-function is the statement of
 existance the unique function depending on infinite set of times and
satisfying
 the infinite hierarchy of equations. What we demonstrated was the only first
 equation of whole hierarchy, see more exhausting consideration in
 \cite{UT,GMMMO}.

 Hereafter, we will not address to any equations and discuss only
 $\tau$-functions.  What we are going to show for the next two sections is that
 {\it any matrix model partition function is a $\tau$-function}. Indeed, using
 (\ref{tauTC}), (\ref{prod}) and putting

 \beq\label{forced}
 \tau_0 = 1,
 \eeq
 one can obtain for one matrix model:

 \beq\label{1}
 {\cal Z}_n = {\tau_n\over \tau_0} = \tau_n.
 \eeq

 \subsubsection{Determinant representation}
 Certainly, to demonstrate that the partition function is a $\tau$-function one
 should not check all infinite hierarchy of equations. During our consideration
 in the next sections, we obtain some effective and convenient representations
 for $\tau$-functions. Now let me just prove that the $\tau$-function of the
 Hermitian one matrix model can be presented in the specific determinant form
 whichj defines, actually, Toda chain $\tau$-function.

 Let us rewrite the orthogonality condition (\ref{OP}) in the "matrix form".
 Namely, we introduce matrix $A$ with matrix elements $a_{mn}$ determined in
 (\ref{OPN}), so-called moment matrix $C$ whose matrix elements are determined
 by

 \beq\label{MomMat}
 C_{ij}= \int dh h^{i+j-2} e^{-V(h)}
 \eeq
 and the diagonal matrix $J$ whose diagonal entries are just $ e^{\varphi _n
}$.
 Then one can rewrite (\ref{OP}) as matrix relation\footnote{This equation is
 nothing but Riemann-Hilbert (factorization) problem - see \cite{UT,GMMMO} for
 details.}  $ACA^T=J$ ($A^T$ means transponed matrix), and taking the
 determinant of both sides of the relation, one obtain

 \beq\label{detrepTC}
 {\cal Z}_n = \det_{n\times n} C_{ij},
 \eeq
 the conditions

 \beq\label{cond1}
 {\partial C_{\ast}(t)\over \partial t_k} \equiv \partial_k  C_{\ast}(t) =
 {\partial^k C_{\ast}(t)\over \partial t_1^k} \equiv \partial^k  C_{\ast}(t),
 \eeq

 \beq\label{cond2}
 C_{ij} = C_{i+j}
 \eeq
 and

 \beq\label{cond3}
 C_n = \partial^{n-2} C_{11} \equiv \partial^{n-2} C
 \eeq
 being satisfied (as it is evident from the manifest form of the moment matrix
 (\ref{MomMat})). Thus, one can write finally

 \beq\label{detrepTC2}
 {\cal Z}_n = \det \partial ^{i+j-2} C.
 \eeq
 Note that the conditions (\ref{cond1}) and (\ref{cond3}) is to be satisfied
for
 any (Toda) KP hierarchy, while (\ref{cond2}) determines Toda chain hierarchy
 \cite{KMMM}. We will return to the question later.

 \subsubsection{$\ast$ Multi-matrix models}
 In this section we discuss the integrability in multi-matrix models
considering
 at the beginning the standard multi-matrix model considered in \cite{MultM}.
 Namely, its partition function is given by the matrix integral\footnote{For
the
 sake of brevity, we omit all proportionality factors, which can be found, say,
 in \cite{Mor}.}

 \beq\label{MMM}
 {\cal Z}^{\{p\}}_n \sim   \int \prod_{i=1}^{p} [dH_i] \exp\left\{-\rm {Tr}
 \left[ \sum_{i=1}^p V_i(H_i) + \sum_{i=1}^{p-1} H_iH_{i-1}\right]\right\}.
 \eeq
 This integral again can be integrated over angular variables, this time using
 Itzykson-Zuber formula \cite{IZ}

 \beq\label{IZ}
 \int [dU]_{n\times n} e^{{\rm tr} XULU^\dagger} \sim
 {{\det _{ij} e^{l_ix_j}}\over {\Delta(L)\Delta(X)}}
 \eeq
 and one obtains

 \ba
 {\cal Z}_n^{\{p\}} \sim  \int \left[\prod_{i=1}^n dh_i^{\{1\}}
 dh_i^{\{p\}}\right] \Delta(h^{\{1\}}) \Delta(h^{\{p\}})
 A(h_i^{\{1\}},h_i^{\{p\}}) \equiv \\ \equiv \int \left[\prod_{l=1}^p
 \prod_{i=1}^n dh_i^{\{l\}}\right] \Delta(h^{\{1\}}) \Delta(h^{\{p\}})
 \exp\left\{- \sum_{i,l} V_l(h_i^{\{l\}}) - \sum_{l=1}^{p-1} \sum_i
 h_i^{\{l\}}h_i^{\{l+1\}}\right\}.\label{eigenMMM}
 \ea
 To apply the orthogonal polynomials in this case, one needs two sets of
 polynomials $Q_n(x)$ and $T_m(y)$, which satisfy the orthogonality relation:

 \beq\label{OPMMM}
 \int dx dy P_n(x)Q_m(y) A(x,y) = \delta_{m,n} h_m(t_k,{\bar t}_k|c_k),
 \eeq
 where times $t_k$ and ${\bar t}_k$ parameterize potentials $V_1$ and $V_p$
 accordingly, while coefficients $c_k$ parameterize other potentials.
 Then, in complete analogy with one matrix model case, one can obtain the
 determinant representation:

 \beq\label{detrepMMM}
 {\cal Z}_n = \det_{n\times n} C_{ij}^{\{p\}}
 \eeq
 where

 \beq\label{MomMatMMM}
 C^{\{p\}}_{ij} \equiv \int dxdy x^{i-1}y^{j-1} A(x,y),\ \ C^{\{p\}}_{11}\equiv
 C^{\{p\}}
 \eeq
 and

 \beq\label{detrepMMM2}
 {\cal Z}_n = \det_{ij} \partial^{i-1}{\bar \partial}^{j-1} C^{\{p\}}
 \eeq
 as the entries in the determinant satisfy the conditions analogous to
 (\ref{cond1}) and (\ref{cond3}), which imply that the partition function
 (\ref{detrepMMM}) is the $\tau$-function of general Toda lattice hierarchy
 \cite{GMMMO,KMMOZ,KMMM}. This property is understood already from the fact
that
 the partition function (\ref{detrepMMM2}) depends essentially on {\it two}
 different sets of times (which correspond to positive and negative times in
 Toda lattice hierarchy), with the other times (potentials) in (\ref{MMM}) just
 parameterizing the $\tau$-function, i.e. the point of the Grassmannian.
 Certainly, it would be more natural to have all times at the equal footing. In
 fact,
 we considered up to now only one possible generalization to multi-matrix
 models. But there is one more, from some viewpoints more attractive,
extension.
 This is so-called "Conformal Multi-Matrix Models" (CMMM) \cite{KMMMP,MP},
which
 will be discussed in the next subsection. These CMMM can be described by
 multi-component hierarchies of $SL(p+1)$ AKNS type and depend on many sets of
 times.

 For the sake of simplicity, we write down here only two matrix model case.
 Then,
 the model is given by the partition function (in eigenvalue variables)

 \beq\label{CMMM}
 {\cal Z}_{n_1,n_2}^{\{CMMM,2\}} \sim \int \left[\prod_{l=1}^2
 \prod_{i_l=1}^{n_l} dh_{i_l}^{\{l\}}\right] \prod_{l=1}^2\Delta^2(h^{\{l\}})
 \prod_{i,j}(h_i^{\{1\}} - h_j^{\{2\}}) \exp\left\{- \sum_{i,l}
 V_l(h_i^{\{l\}})\right\} .
 \eeq
 We will discuss later why this quite specific partition function is especially
 interesting and discuss its integrable structure in more details. Now it is
 just an example of another multi-matrix model which has a rather simple
 determinant representation

 \ba
 {\cal Z}_{n_1,n_2}^{\{CMMM,2\}} =\\= \det \left[{
 \begin{array}{cccccc}
 C&\ldots&\partial^{n_1-1}C&
 \bar C&\ldots&
 \bar\partial^{n_2-1}\bar C\\
 \partial C&\ldots&\partial^{n_1}C&
 \bar\partial \bar C&\ldots&
 \bar\partial^{n_2}\bar C\\
 \vdots&\ddots&\vdots&\vdots&\ddots&\vdots\\
 \partial^{n_1+n_2-1}C&\ldots&\partial^{2n_1+n_2-2}C&
 \bar\partial^{n_1+n_2-1}\bar C&\ldots&
 \bar\partial^{2n_1+n_2-2}\bar C
 \end{array}}\right]\label{CMMMdetrep}
 \ea
 where

 \beq
 C(t) = \int   dz\ \exp  [-V_1(z)],  \ \ \    \bar C(\bar t) = \int   dz\ \exp
 [- V_2(z)].
 \eeq
 This representation gives rise to the $\tau$-function of (generalized) $SL(3)$
 AKNS system \cite{Newell,KMMMP}. From this view point, one matrix model
 corresponds to usual $SL(2)$ AKNS hierarchy, which is equivalent to Toda chain
 \cite{Newell}. Certainly, analogous equivalence is absent for higher
 $SL(N)$ hierarchies.

 \subsection{Virasoro constraints}
 \setcounter{equation}{0}
 \subsubsection{Ward identities in matrix models}
 In the previous subsection we briefly described the integrable properties of
 the discrete matrix models. In this subsection we pay attention to equally
 important and general property, constraint algebra which is imposed on matrix
 models. The role of these constraint algebra is to fix the concrete
 $\tau$-function. In fact, the integrable hierarchy does not fix the
 $\tau$-function uniquely. Moreover, there is a huge space of the solution to
 the equations of hierarchy, which are parameterized by (subspaces) of infinite
 dimensional Grassmannian. It turns out that constraint algebra imposed on
 $\tau$-function fixes it usually uniquely. It means that in order to describe
 matrix model parittion function in invariant way, one should determine what
 kind of $\tau$-function is the corresponding partition function (what
hierarchy
 it is described) and what constraint algebra it satisfies. Again, let us start
 with the simplest example of Hermitian one matrix model.

 We demonstrate now that the partition function (\ref{1MM}) satisfies Virasoro
 algebra constraints (more precisely, its Borel subalgebra). Indeed, these are
 nothing but Ward identities of the model (\ref{1MM}) \cite{MM,AMJ,IM,D}. To
see
 this, let us shift the variable $H$ by small quantity $\epsilon H^{q+1}$.
Then,
 as it is nothing but the change of variables, the integral (\ref{1MM}) will be
 unchanged. From the other hand, this shift results into changing of potential
 $\sum t_kH^k \longrightarrow \sum_k (H+\epsilon H^{q+1})^k = \sum_{k=1}^q
 t_kH^k + \sum_{k>q} (t_k + (k-q)t_{k-q})H^k)$. This means that these shifts
are
 given by the action of operators $L_q^{cl} \equiv \sum kt_k {\partial\over
 \partial t_{k+q}}$ producing Virasoro algebra.

 Now we should also take into account the changing of the measure under the
 transformation above. The Jacobian of variation of Haar measure on Hermitian
 matrices is equal to $\det \|{\partial(H_{ij}+\epsilon H^{q+1})_{ij}\over
 \partial H_{kl}} \|\sim 1 + \epsilon \Tr {\partial H^{q+1}\over \partial H} =
1
 + \sum_{k=0}^q {\rm Tr} H^k {\rm \Tr} H^{q-k}$. As any degree $k$ of matrix
$H$
 can be obtained by differentiating of the exponential of potential with
respect
 to $t_k$, one can reproduce the changing of measure by the operators
 $\sum_{k=0}^q {\partial^2\over \partial t_k\partial t_{q-k}}$. Note that
 according to the definition (\ref{1MM}) $$\frac{\partial}{\partial t_0}{\cal
 Z}_n = n{\cal Z}_n.$$ Therefore, the final result of the shift is:

 \beq\label{WI}
 {\cal Z}_n(t)= {\cal Z}_n(t)\left|_{{\rm with\ shifted\ H}} = (1+L_q){\cal
 Z}_n(t),\right.
 \eeq
 i.e.

 \beq\label{VirWI}
 L_q{\cal Z}_n = 0, \ \ q\ge -1,
 \eeq
 where operators

 \beq\label{Virasoro}
 L_q = \sum_{k=0}^\infty kt_k {\partial \over \partial t_{k+q}} + \sum_{k=0}^q
 {\partial^2 \over \partial t_k \partial t_{q-k}}
 \eeq
 satisfy the commutation relations of Virasoro algebra

 \beq\label{VirComm}
 \phantom{ghfgh}[L_n,L_m]=(n-m)L_{n+m}.
 \eeq
 As we consider only Borel subalgebra of whole Virasoro algebra, the question
of
 the central charge is out of our consideration.

 Thus, we proved that the partition function of Hermitian one matrix model
 satisfies Virasoro algebra constraints. Unfortunately, it is not so simple to
 find out an analogous symmetry in multi-matrix models (\ref{MMM}). So, we are
 going to suggest some more general way to construct matrix models possessing
 given symmetry, which allows one to write down multi-matrix models with
 $W$-symmetry, and, as a by-product, explains why conformal algebra arises in
 matrix models.

 \subsubsection{Conformal matrix models}
 In fact, now we just demonstrate that the partition function of (\ref{1MM})
can
 be rewritten in terms of correlators in gaussian $c=1$ confomal field theory
 and  the Virasoro generators (\ref{Virasoro})
 actually have the well-known form of the Virasoro operators in
 the theory of one free scalar field.
 Indeed, let us consider
 {\it holomorphic} components of the scalar field
 \bea\label{freesc}
 \phi (z) &=  \hat Q + \hat P \log z  + \sum _{k\neq 0} {J_{-k}\over k}
 z^{-k}\nn\\
 \  [J_n,J_m] &= n\delta _{n+m,0},  \ \ \     [\hat Q,\hat P] = 1
 \eea
 and define the vacuum states
 \bea\label{bosvac}
 J_k|0\rangle  &= 0, \ \ \  \langle n|J_{-k} = 0, \ \ \    k > 0\nn\\
 \hat P|0\rangle  &= 0, \ \ \   \langle N|\hat P = n\langle n|.
 \eea
 ''Half" of the stress-tensor\footnote
 {For the sake of brevity, we omit the sign of normal ordering in the
 evident places, say, in the expression for $T$ and $W$ in terms of
 free fields.} components
 \beq\label{tensor}
 T(z) = {1\over 2}[\partial \phi (z)]^2 = \sum    T_qz^{-q-2},\quad
 T_q = {1\over 2}\sum _{k>0}J_{-k}J_{k+q} +
 {1\over 2}\sum _{{a+b=q}\atop{a,b\geq 0}}J_aJ_b,
 \eeq
 obviously vanishes the $SL(2)$-invariant vacuum
 \beq\label{6a}
  T_q|0\rangle  = 0,  \ \ \    q \geq  -1.
 \eeq
 Then, we define the Hamiltonian by
 \bea\label{ham}
 H(t) &= {1\over \sqrt{2}} \sum _{k>0}t_kJ_k =
 \oint_{C_0}V(z)j(z)\nn\\
 V(z) &= \sum _{k>0}t_kz^k, \ \  \   j(z) = {1\over \sqrt{2}}\partial \phi (z).
 \eea
 Now let us solve the Virasoro constraints (\ref{VirWI}) in the general form.
 Put differently, we derive the general parition function possessing the
 Virasoro symmetry (\ref{VirWI}). One can easily construct a ``conformal field
 theory" solution of (\ref{VirWI}) in two steps. The basic ''transformation"
 \beq\label{trick}
 L_q\langle n|e^{H(t)}\ldots = \langle n|e^{H(t)}T_q\ldots
 \eeq
 can be checked explicitly and show how the Virasoro generators
(\ref{Virasoro})
 transforms to those in gaussian model. As an immediate consequence, any
 correlator of the
 form
 \beq\label{dcor}
 \langle n|e^{H(t)}G|0\rangle
 \eeq
 ($n$  counts the number of zero modes, ''included" in  $G$ -- that is the
 role of the {\it size} of matrix in (\ref{1MM})) gives a
 solution to
 (\ref{VirWI})
 provided by
 \beq\label{com}
 [T_q,G] = 0, \ \ \  q \geq  -1.
 \eeq
 The conformal solution to (\ref{com})
 (and therefore to (\ref{WI})) immediately comes from the basic
 properties of $2d$ conformal algebra. Indeed, any solution to
 \beq\label{comm}
 [T(z),G] = 0
 \eeq
 is a solution to (\ref{com}),
 and it is well-known that the solution to (\ref{comm}) is (by definition
 of the chiral algebra) a
 function of {\it screening charges} in the free scalar field theory
 given by
 \beq\label{scr}
 Q_\pm  = \oint J_\pm  = \oint
 e^{\pm \sqrt{2}\phi }.
 \eeq
 With a selection rule on zero mode it gives\footnote{Of course, the general
 case might be  $G \sim  Q^{n+m}_+Q^m_-$ but the special
 prescription for integration contours, proposed in \cite{MMM},
 implies that the
 dependence of $m$ can be irrelevant and one can just put  $m = 0$. In fact,
the
 problem with the choice of integration contour is a rather subtle point and
 discussed in \cite{MMM}.}
 \beq\label{discrG}
 G = \exp \ Q_+ \rightarrow  {1\over n!}Q^n_+
 \eeq
 In this case the solution
 \beq\label{z2cor}
 {\cal Z}_n(t) = \langle n|e^{H(t)}\exp Q_+|0\rangle
 \eeq
 after computation of the free theory correlator, analytic continuation of the
 integration contour
 gives the result
 \ba\label{z2}
 {\cal Z}_n = (n!)^{-1}\int   \prod ^n_{i=1}dz_i \exp  \left( - \sum
 t_kz^k_i\right)  \Delta ^2_n(z) =\\
 = (n!{\rm Vol}\ U(n))^{-1}\int   [dH]\ \exp \left( - \sum \Tr   t_kH^k\right)
 \ea
 in the form of the matrix integral (\ref{1MM}). This is the simplest example
of
 CMMM mentioned above which already explains the name.

 Thus, we derive Virasoro invariant matrix model of the general type.
Certainly,
 this procedure is immediately generalized to other symmetries.

 \subsubsection{$\ast$ Ward identities in multi-matrix models}
 Now we are going to derive multi-matrix models invariant with respect to
 $W$-algebra and discuss what is the symmetry of the standard multi-matrix
 models (\ref{MMM}).

 Let us consider the immediate generalization of proposed procedure. Now we
 can use powerful tools of $2d$ conformal field theories, where it is well
 known how to
 generalize almost all the steps of above construction: first, instead of
 looking for a solution to Virasoro constraints one can impose {\it extended
 Virasoro} or  $W$-constraints on the partition function. In such case one
would
 get Hamiltonians in terms of {\it multi}-scalar field theory, and the second
 step is generalized directly using {\it screening charges} for  $W$-algebras.
 The general scheme of solving discrete $W$-constraints looks as
 follows:

 (i)  Consider Hamiltonian as a linear combination of the Cartan currents of a
 level one Kac-Moody algebra  ${\cal G}$
 \beq\label{hamil}
 H(t^{(1)},\ldots,t^{({\rm rank}\ {\cal G})}) =
 \sum _{\lambda ,k>0}t^{(\lambda )}_k\bmu _\lambda \bJ_k,
 \eeq
 where $\{\bmu_i\}$ are basis vectors in Cartan hyperplane, which
 for $SL(p+1)$ case are chosen to satisfy
 $$
 \bmu_i\cdot \bmu_j=\delta_{ij}-{1\over p+1},  \ \ \ \sum_{j=1}^{p+1}
 \bmu_j=0.
 $$

 (ii)  The action of differential operators  ${\cal W}^{(a)}_i$ with respect to
 times  $\{t^{(\lambda )}_k\}$ can be now defined from the relation
 \beq\label{wtrick}
 W^{(a)}_i\langle \bn|e^{H(\{t\})}\ldots =
 \langle \bn|e^{H(\{t\})}{\rm W}^{(a)}_i\ldots\ , \ \ \   a=2,\ldots,p+1;  \ \
\
 i\geq 1-a,
 \eeq
 where
 \ba\label{wmodes}
 \W^{(a)}_i = \oint z^{a+i-1}\W^{(a)}(z)\\
 \W^{(a)}(z) = \sum  _\lambda  [\bmu _\lambda \partial \bphi (z)]^a + \ldots
 \ea
 are  spin-$a$ W-generators of  $\W_{p+1}$-algebra written in terms of
 rank$\;{\cal G}$-component scalar fields \cite{FL}.

 (iii)  The conformal solution to the discrete $W$-constraints arises in the
 form
 \beq\label{zcmm}
 {\cal Z}^{\{CMMM,p+1\}}_{\bsn}(t) = \langle \bn|e^{H(\{t\})}G\{Q
 ^{(\alpha)} \}|0\rangle
 \eeq
 where  $G$  is again an exponential function of screenings of level one
 Kac-Moody algebra (see \cite{GMMOS} and references therein)
 \beq\label{screen}
 Q^{(\alpha)}  = \oint J^{(\alpha)}  = \oint e^{\bsalpha \bsphi }
 \eeq
 $\{\balpha \}$ being roots of finite-dimensional simply laced
 Lie algebra ${\cal G}$.
 The correlator (\ref{zcmm}) is still a free-field correlator
 and the computation gives it again in a multiple integral form
 \ba\label{multi}
 {\cal Z}^{\{CMMM,p+1\}}_{\bsn}(t) \sim  \int   \prod  _\alpha
 \left[ \prod ^{n_\alpha }_{i=1}dz^{(\alpha )}_i \exp \left( -
 \sum _{\lambda ,k>0}t^{(\lambda )}_k(\bmu _\lambda \balpha )(z^{(\alpha
)}_i)^k
 \right) \right] \times \\
 \times \prod _{(\alpha ,\beta )}\prod ^{n_\alpha }_{i=1}
 \prod ^{n_\beta }_{j=1}(z^{(\alpha )}_i- z^{(\beta )}_j)^{\bsalpha \bsbeta }.
 \ea
 The only difference with the one-matrix case (\ref{z2}))
  is that the
 expressions (\ref{multi}) have rather complicated representation in terms of
 multi-matrix integrals, the following objects will necessarily appear
 \ba\label{matten}
 \prod ^{n_\alpha }_{i=1} \prod ^{n_\beta }_{j=1}(z^{(\alpha )}_i-
 z^{(\beta )}_j)^{\bsalpha \bbeta } \longrightarrow \left[ \det \{H^{(\alpha
 )}\otimes I -
 I\otimes H^{(\beta )}\}\right] ^{\bsalpha \bsbeta },
 \ea
 However, this is still a model with a chain of matrices
 and with closest neighbour interactions only (in the case of $SL(p+1)$).

 Actually, it can be shown that CMMM, defined by (\ref{zcmm}) as a
 solution to the $W$-constraints has a very rich integrable structure
 and possesses a natural continuum limit \cite{KMMMP,MP}. To pay for these
 advantages one
 should accept a slightly less elegant matrix integral with the entries like
  (\ref{matten}).

 The first non-trivial example is the $p=3$ solution to
 $W_3$-algebra: an alternative to the conventional 2-matrix model.
 In this case
 one has six screening charges  $Q^{(\pm \alpha _i)}$ $(i = 1,2,3)$
 which commute with
 \beq\label{w2}
 \W^{(2)}(z) = T(z) = {1\over 2}[\partial \bphi (z)]^2
 \eeq
 and
 \beq\label{w3}
 \W^{(3)}(z) = \sum ^3_{\lambda =1}(\bmu _\lambda \partial \bphi (z))^3,
 \eeq
 where  $\bmu _\lambda $ are vectors of one of the fundamental representations
 ({\bf 3} or $\bar {\bf 3})$ of  $SL(3)$.

 The particular form of integral representation (\ref{multi})
 depends on particular screening insertions to the correlator (\ref{zcmm}).
 We will concentrate on the solutions which have
 no denominators. One of the reasons of
 such choice is that these solutions possess the most simple integrable
 structure,
 though the other ones can still be analyzed in the same manner.

 The simplest solutions which have no denominators correspond to specific
 correlators
 \beq
 {\cal Z}^{\{CMMM,p+1\}}_{\bsn}(t) = \langle \bn|e^{H(\{t\})}\prod_i
 \exp Q_{\alpha _i}|0\rangle
 \eeq
 when we take  $\alpha _i$ to be ``neighbour" (not simple!) roots:
 $(\balpha _i\balpha _j) =
 1$. In the case of  $SL(3)$  this corresponds, say, to  insertions of only
    $Q_{\alpha _1}$ and  $Q_{\alpha _2}$ and gives formula (\ref{CMMM}).
 Thus, we establish that the partition function (\ref{CMMM}) satisfies
 $W^{\{3\}}$-constraint algebra, the generators of which can be read off from
 the formula (\ref{wtrick}):

 \beq\label{W3}
 {W}^{\{3\}}_q =
  3 \sum _{k,l>0}kt_klt_l {\partial \over \partial t_{q+k+l}} +
 3 \sum _{k>0}kt_k\sum _{a+b=k+q}{\partial ^2\over \partial t_a
 \partial t_b} +
 \sum _{a+b+c=q}{\partial ^3\over \partial t_a\partial t_b\partial t_c}.
 \eeq

 There are many possible generalizations of the above scheme, some of them,
 including ''supersymmetric matrix models", can be found in \cite{KMMMP}.

 Thus, we derived matrix models possessing $W$-symmetry. The question we are
 going to discuss to complete this subsection is what symmetry does possess the
 standard multi-matrix model (\ref{MMM}). Let us consider the simplest case of
 two-matrix model (\ref{MMM}), one of the potentials, say, $V_2$ being
truncated
 to be just $cH_2^3$. Now we repeat the procedure of the derivation of the
 Virasoro constraints (\ref{WI}). Let us do the following infinitesimal change
 of integration
 variables

 \ba\label{G1}
 \delta H_2  =  H_1 ^q\hbox{, }    q \geq  0 \\
 \delta H_1  =  3c H_1^q \left( V_1'(H_1) -  H_2\right) + \hbox{"quantum    }
 \hbox{  corrections"}.
 \ea
 This variation of variables induces the variation of potential $V_1$:

 \beq\label{G2}
 \delta V_1 = -3c(V')^2 H_1^q + H_1^{q+1} + \hbox{"quantum   }
 \hbox{  corrections"}.
 \eeq
 The first term in this expression gives rise to the "classical" part
 of the symmetry generators (analogous to  $L_q^{cl}$ above) and the second one
 produces the derivative
 $\partial / {\partial t_{q+1}}$ in the Ward identities.

 While the $H_2$-component of the variation (\ref{G1}) does not change the
 integration measure  $dH_1dH_2$, this is not
 true for  $H_2$-component. The corresponding Jacobian is responsible for
 the ``quantum" contributions to (\ref{G1}) and (\ref{G2}).

 Now we just write down the result for the constraints imposed on the partition
 function (\ref{CMMM}):

 \beq\label{GavaWI}
 \left.\left( - {\partial \over \partial t_{q+1}} + 3c\tilde
 {W}^{\{3\}}_{q-2}\{t\}\right) {\cal Z}_n^{\{2\}}(t)\right|_{V_2=cH_2^3} = 0.
 \eeq
 where generators

 \ba\label{G3}
 \tilde {W}^{\{3\}}_{q-2} = \sum _{k,l>0}kt_klt_l
 {\partial \over \partial t_{q+k+l-2}} +
 \sum _{k>0}kt_k\sum _{a+b=k+q-2}{\partial ^2\over \partial t_{a}\partial t_b}
+
 \\
 +
 \sum _{k>0}kt_k\sum _{a+b=q-1}{\partial ^2\over \partial t_a\partial t
 _{k+b-1}} +
  \sum _{a+b+c=q-2}{\partial ^3\over \partial t_a\partial t_b\partial t_c} +
 {q(q-1) \over 2} {\partial \over \partial t_q},
 \ea
 produces some new constraint algebra.

  This $\tilde {W}^{\{3\}}$ algebra arises in many different contexts and can
be
 trivially generalized to any spin.
 The generators of this algebra have the same "classical part" as the standard
 $W^{\{3\}}$ algebra (\ref{W3}), but "quantum parts" are different. The set of
 $\tilde W$-constraints is closed (as well Ward identities (\ref{GavaWI})) and
 is an algebra quadratic in generators, similar to the standard $W$-algebra.
Let
 us stress that considering the polynomial potential $V_2$ of a degree $K$, one
 obtains the Ward identity, analogous (\ref{GavaWI}), but for $\tilde
 W^{\{K\}}$-algebra. In particular, the generic potential leads to linear
 ${\tilde W}^{\{\infty\}}$-algebra. The detailed description of the properties
 of $\tilde W$-algebra and constraints in two-matrix model (\ref{MMM}) can be
 found in \cite{Gava,Mor}.

 Thus, we realized that the simple multi-matrix models (\ref{MMM}) satisfy
 unusual $\tilde W$-constraints, while more complicated models (\ref{CMMM})
 satisfy the standard $W$-constraints. In the continnum limit this difference
 disappears, and both these types of models have the same (double scaling)
 continuum limit. Unfortunately, we do not know the efficient way of taking the
 continuum limit in the multi-matrix models (\ref{MMM}) (certainly, I mean by
 this that we have no complete understanding of the renormalization of all
 operators and constraint algebra, i.e. of simultaneously all correlation
 functions). But CMMM turns out to be more suitable for the continuum limit
 procedure. This is one of their advantages. We are going to discuss these
 questions in the next subsection.

 \subsection{Continuum limit}
 \setcounter{equation}{0}
 \subsubsection{Continuum limit of Toda chain equations}
 In this subsection we duscuss the bridge between above considered discrete
 matrix models and these are in the double scaling limit. Let us stress that,
 due to singularity of this procedure, it is very difficult to take this
 continuum limit directly in the partition function (in particular, there
should
 be some (infinite) renormalizations), but easier to do in some equations,
where
 the porcedure is less singular (to be honest, there is a very simple way of
 doing the double scaling limit immediately in the partition function, but it
 requires some additional information and will be discussed in the next
 section). Indeed, we already mentioned that the crucial properties of
 integrability and imposed constraint algebra are preserved in the continuum
 limit. It implies that the most natural choice of equations to take the
 continuum limit would be either constraint algebra, or integrable equations.
 Now we work out both these ways for the simplest example of the Hermitian one
 matrix model (\ref{1MM}).

 Let us start with integrable equations of Toda chain hierarchy. Our further
 discussion is sufficiently general and does not refer to the concrete matrix
 model $\tau$-function, i.e. the continuum limit conserves all the space of
 solutions to integrable equations, and the structure of integrability. From
the
 other hand, we demonstrate below that it also conserves constraint algebra. It
 is specific of the double scaling limit, which can be invariantly determined
 just by its properties of not disturbing all crucial matrix model properties

 Thus, now we build the map from Toda hierarchy to KdV hierarchy by limiting
 procedure, and then we do analogous map from discrete Virasoro algebra
 (\ref{Virasoro}) to that in the double scaling limit. Certainly, we consider
 here only the first equation of the hierarchy (\ref{TodaEq}). For the sake of
 simplicity, we consider here only the case of even potential in (\ref{1MM}).
It
 implies that all odd times $t_{2k+1}=0$ and all $\varphi_n$ are supposed to be
 independent of them. Thus, $p_n={\partial \varphi_n\over \partial t_1} = 0$
and
 the first non-trivial equation (which is really the second equation of Toda
 chain hierarchy) has a form ($t\equiv t_2$):

 \beq\label{Volterra}
 \frac{\partial R_n}{\partial t} = - R_n(R_{n+1}-R_{n-1}).
 \eeq
 This hierarchy is the reduction of Toda chain hierarchy and called Volterra
 hierarchy.

 Indeed, we could guess that the continuum limit of Toda chain hierarchy should
 be KdV hierarchy by looking at the Lax operator (\ref{Loper}). Indeed, the
main
 idea of the continuum limit in matrix models is to change the discrete
variable
 $n$ by the continuous time variable (which will be the first time in the
 continuum hierarchy). Therefore, choosing $n=x\epsilon$ we can get in the
 continuum limit:

 \beq\label{3}
 {\rm {\bf L}}^{cont} = 2 - \epsilon^2 [\partial ^2_x + (\partial \varphi
(x))].
 \eeq
 One can recognize in this operator just Lax operator of KdV hierarchy, the
 eigenvalues of Toda Lax operator being distributed around zero in the
continuum
 limit (see \cite{GMMMO}). Still for the equations the continuum limit
procedure
 is not so simple. Indeed, doing naively, we get from (\ref{Volterra}):

 \beq\label{DispLess}
 \frac{\partial R(x)}{\partial t} = -
 R(x)(R(x+\epsilon)-R(x-\epsilon))\stackreb{\epsilon \to 0}{=} -
 R(x')R'(x').
 \eeq
 provided by the replace $x=\epsilon x'$. This formula is called dispersionless
 KdV equation (or Bateman, or Hopf, or Khohlov-Zabolotskaya equation). This
 equation is too simple and corresponds to just naive "large $n$ limit" of
 matrix models \cite{BIPZ}. To obtain KdV equation, one should take care of
 non-linear term. The procedure allowing one to preserve such terms is called
 "double scaling limit" and is as follows.

 Imagine, that in
 continuum limit $R_n$ tends to a constant $R_0$, and the function $r(x)$
 arises only as scaling approximation to this constant:
 $R(x) = R_0(1 + \epsilon ^sr(x))$. Then the leading term at the r.h.s. of
 (\ref{DispLess}) is $\epsilon RR'(x) =
 -2\epsilon^s r(x) (1 + {\cal O}(\epsilon^2,\epsilon^s))$,
 and instead of (\ref{DispLess}) we would get:

 \beq\label{4}
 \frac{\partial r}{\partial t} = - 2\epsilon R_0 r'(x) ((1 + {\cal
 O}(\epsilon^2,\epsilon^s)).
 \eeq
 Then, by a simple change of
 variables\footnote{
 This change of variables is implied by the relation:

 \beq
 \frac{\partial}{\partial t} + 2\epsilon^sR_0 \frac{\partial}{\partial x} =
 \left( \frac{\partial \tilde t}{\partial t} +
 2\epsilon^sR_0 \frac{\partial \tilde t}{\partial x}\right)
 \frac{\partial}{\partial \tilde t} +
 \left( \frac{\partial \tilde x}{\partial t} +
 2\epsilon^sR_0 \frac{\partial \tilde x}{\partial x}\right)
 \frac{\partial}{\partial \tilde x} =
 \frac{\partial}{\partial \tilde t}.
 \eeq
 }

 \beq
 T_1 = x - 2\epsilon R_0 t, \\
 T_3 = \epsilon^3 R_0 t
 \label{chavacont}
 \eeq
 it can be transformed into

 \beq\label{6}
 \frac{\partial r}{\partial T_3} = \epsilon^{-2} {\cal
 O}(\epsilon^2,\epsilon^s),
 \eeq
 and terms at the r.h.s. also should be taken into account.
 Then we get:

 \ba\label{5}
 \frac{\partial r(x)}{\partial t} = -2\epsilon R_0 \left(1 + \epsilon^sr(x))
 (r'(x) + \frac{1}{6}\epsilon^2r'''(x) + {\cal O}(\epsilon^4)\right) = \\
 = -2\epsilon R_0 \left( r'(x) + \frac{1}{6}\epsilon^2r'''(x)
 + \epsilon^s rr'(x) + \epsilon^2{\cal O}(\epsilon^2, \epsilon^s)\right)
 \ea
 and, after the change of variables (\ref{chavacont}),

 \beq
 \frac{\partial r(T_1)}{\partial T_3} = - \frac{1}{3} r'''(T_1)
 - 2\epsilon^{s-2}rr'(T_1) + {\cal O}(\epsilon^2, \epsilon^s).
 \eeq
 One can see that the choice $s=2$  is distinguished (a critical point)
 and at this point we get nothing but KdV equation:

 \beq\label{KdV}
 \frac{\partial r}{\partial T_3} =
 -\frac{1}{3}\frac{\partial^3r}{\partial T_1^3} -
 2r\frac{\partial r}{\partial T_1},
 \eeq

 Thus, we obtained KdV equation. Certainly, one can get all KdV hierarchy in
the
 same way. Moreover, one can investigate the continuum limit of the Toda (not
 Volterra) hierarchy, producing {\it two equivalent} KdV hierarchies (which
 correspond to flows in odd and even Toda times) - see \cite{GMMMMO,BP}. It
 reflects in some problems in the continuum limit of Virasoro algebra (see
 below).

 Thus, the lesson, which we can extract from this continuum limit procedure is
 that, to get the proper limit, one should do linear transformation of times
and
 renormalization of the partition function (or $R_n$). Now we use these lessons
 in order to study the continuum limit of Virasoro algebra.

 \subsubsection{$\ast$ The continuum limit of the Virasoro algebra}
 The procedure of taking the continuum limit of Virasoro algebra we demonstrate
 now is a bit tedious and too technical, as, in contrast to our previous
 consideration allows one to work with whole hierarchy simultaneously. Let me
 just describe a set of rules which leads to the correct answer, all details
and
 hints can be found in \cite{MMMM}.

 It has been suggested in \cite{FKN} that the square root of the
 partition function of the continuum limit of
 one-\mm is subjected to the Virasoro constraints

 \beq\label{b1}
 {\cal L}_q^{\rm cont}\sqrt{{\cal Z}^{\rm ds}}=0,\quad q\geq -1,
 \eeq
 where

 \ba \label{b2}
 {\cal L}^{\rm cont}_q=\sum_{k=0}\left(k+{1\over2}\right)
 T_{2k+1}{\partial\over\partial T_{2(k+q)+1}}+G
 \sum_{0\leq k\leq q-1}{\partial^2\over\partial T_{2k+1}\partial
 T_{2(q-k-1)+1}}+\\
 +{\delta_{0,q}\over16} +{\delta_{-1,q}T_1^2\over(16G)}
 \ea
 are modes of the stress tensor

 \beq\label{b3}
 {\cal T}(z)= {1\over 2}{:}\partial\Phi^{2}(z){:} - {1\over16z^2}
 =\sum{{\cal L}_q\over z^{q+2}}.
 \eeq

 Indeed, we demonstrate that these equations which reflect the
 $W^{\{2\}}$-invariance of the partition function of the continuum model can
 be deduced from analogous constraints in Hermitian one-\mm by taking the
 double-scaling continuum limit.
 The procedure is as follows.

 In order to obtain the relation between $W$-invariance of the
 discrete and continuum models one has to consider a reduction of  model
 (\ref{1MM}) to the pure even potential $t_{2k+1}=0$.
 Let us denote by the $\tau_n^{\rm red}$ the partition function of the
 reduced \mm

 \beq    \label{a1}
 \tau^{\rm red}_{n}\{t_{2k}\}=\int [dH]\exp{\rm Tr}\sum_{k=0}t_{2k}
 H^{2k}
 \eeq
 and consider the following change of the time variables

 \beq\label{a5}
 g_m=\sum_{k\geq m}{(-)^{k-m}\Gamma\left(k+{3\over2}\right)
 a^{-k-{1\over2}}\over(k-m)!\Gamma\left(m+{1\over2}\right)}T_{2k+1},
 \eeq
 where $g_m \equiv mt_{2m}$ and this expression can be used also for the zero
 discrete time $g_0 \equiv n$ that plays the role of the dimension of matrices
 in the one-matrix model. Derivatives with respect to  $t_{2k}$ transform as

 \beq\label{a6}
 {\partial\over\partial t_{2k}}=\sum_{m=0
 }^{k-1}{\Gamma\left(k+{1\over2}\right)
 a^{m+{1\over2}}\over(k-m-1)!\Gamma\left(m+{3\over2}\right)}{\partial
 \over\partial \tilde T_{2m+1}},
 \eeq
 where the auxiliary continuum times $\tilde T_{2m+1}$ are connected with
 ``true'' Kazakov continuum times $T_{2m+1}$ via

 \beq\label{a7}
 T_{2k+1}=\tilde T_{2k+1}+a{k\over k+1/2}\tilde T_{2(k-1)+1},
 \eeq
 and coincide with $T_{2m+1}$ in the double-scaling limit when $a\to0$.

 Let us rescale the partition function of the reduced one-\mm by
 exponent of quadratic form of the auxiliary times $\tilde T_{2m+1}$

 \beq\label{a11}
 \tilde\tau=\exp\left(-{1\over2}\sum_{m,k\geq0}A_{mk}\tilde T_{2m+1}
 \tilde T_{2k+1} \right)\tau^{\rm red}_n
 \eeq
 with

 \beq\label{a12}
 A_{km}={\Gamma\left(k+{3\over2}\right)\Gamma\left(m+{3\over2}\right)\over
 2\Gamma^2\left({1\over2}\right)}
 {(-)^{k+m}a^{-k-m-1}\over k!m!(k+m+1)(k+m+2)}.
 \eeq
 Then a direct though tedious calculation \cite{MMMM} demonstrates that
 the relation

 \beq\label{a15}
 {\tilde{\cal L}_q\tilde\tau\over\tilde\tau}
 =a^{-q}\sum_{p=0}^{q+1}C^p_{q+1}(-1)^{q+1-p}
 {L_{2p}^{\rm red}\tau^{\rm red}\over \tau^{\rm red}},
 \eeq
 is valid, where

 \beq
 L_{2q}^{\rm red} \equiv \sum_{k=0}kt_{2k}{\partial\over\partial t_{2(k+q)}}+
 \sum_{0\leq k\leq q}{\partial ^2\over \partial t_{2k}\partial t_{2(q-k)}}
 \eeq
 and

 \ba\label{b6}
 \tilde{\cal L}_{-1}=\sum_{k\geq1}\left(k+{1\over2}\right)T_{2k+1}
 {\partial\over\partial \tilde T_{2(k-1)+1}}+
 {T_1^2\over16G},\\
 \tilde{\cal L}_{0}=\sum_{k\geq0}\left(k+{1\over2}\right)T_{2k+1}
 {\partial\over\partial \tilde T_{2k+1}},\\
 \tilde{\cal L}_q=\sum_{k\geq0}\left(k+{1\over2}\right)T_{2k+1}
 {\partial\over\partial \tilde T_{2(k+q)+1}} \\
 +\sum_{0\leq k \leq q-1}{\partial\over\partial \tilde T_{2k+1}}
 {\partial\over\partial \tilde T_{2(q-k-1)+1}} -{(-)^q\over16a^q},\ \ \
 n\geq1.
 \ea
 Here $C^p_q =\frac{q!}{p!(q-p)!}$ are binomial coefficients.

 These Virasoro generators differ from the Virasoro generators
 (\ref{b2})  \cite{FKN,DVV}
 by terms which are singular in the limit $a\longrightarrow 0$.
 At the same time $L_{2p}^{\rm red}\tau^{\rm red}$ at the r.h.s. of (\ref{a15})
 do
 not need to vanish, since

 \beq\label{15aa}
 0 = L_{2p}\tau\left\vert_{t_{2k+1}=0} =
 L_{2p}^{\rm red}\tau^{\rm red} +
 \sum_i {\partial^2\tau\over\partial t_{2i+1}\partial t_{2(p-i-1)+1}}
 \right\vert_{t_{2k+1}=0}.
 \eeq
 It was shown in \cite{MMMM} that these two origins of difference between
 (\ref{b2}) and (\ref{b6}) actually cancel each other, provided eq.(\ref{a15})
 is rewritten in terms of the square root $\sqrt{\tilde\tau}$ rather than
 $\tilde\tau$ itself:

 \beq\label{a16}
 {{\cal L}^{\rm cont}_q\sqrt{\tilde\tau}\over\sqrt{\tilde\tau}}
 =a^{-q}\sum_{p=0}^{q+1}C^p_{q+1}(-1)^{q+1-p}\left. {L_{2p}\tau\over\tau}
 \right\vert_{t_{2k+1}=0}\left( 1+  O(a) \right).
 \eeq
 The proof of this cancelation, as given in  \cite{MMMM}, is not too much
 simple and makes use of integrable equations for $\tau$.

 \subsubsection{$\ast$ Invariant formulation of the limiting procedure}
 After we demonstrated the continuum limit procedure for the Hermitian one
 matrix case, the next question to be addressed to is how to generalize this
 limiting procedure to multi-matrix case. Indeed, in the case of conventional
 multi-matrix models this procedure is still unknown. But for CMMM it can be
 trivially done, what is one of the advantages of CMMM we mentioned above. To
do
 this, let us demonstrate a more economical way to
 define the change of the time-variables $t \longrightarrow T$
 (also discussed in \cite{MMMM}), implied by the scalar field formalism.
 The Kazakov change of the time variables (\ref{a5},\ref{a6}) can be
 deduced from the following prescription. Let us consider the free
  scalar field with periodic boundary conditions

 \beq\label{a8}
 \partial\varphi(u)=\sum_{k\geq0}g_ku^{2k-1}+
 \sum_{k\geq1}{\partial\over\partial t_{2k}}u^{-2k-1},
 \eeq
 and analogous scalar field with antiperiodic boundary conditions:

 \beq\label{a9}
  \partial\Phi(z)=\sum_{k\geq0}\left(\left(k+{1\over2}\right)T_{2k+1}
 z^{k-{1\over2}}+
 {\partial\over\partial \tilde T_{2k+1}}z^{-k-{3\over2}}\right).
 \eeq
 Then the equation

 \beq\label{a10}
 {1\over\tilde\tau}\partial\Phi(z)\tilde\tau =
 a {1\over\tau^{\rm red}}\partial\varphi(u)\tau^{\rm red},\quad
 u^2=1+az
 \eeq
 generates the correct transformation rules
 (\ref{a5}), (\ref{a6}) and gives rise to the expression
 for $A_{km}$ (\ref{a12}).
 Taking  the square of the both sides of the identity (\ref{a10}),

 \bea\label{a100}
 &{1\over\tilde\tau}{\cal T}(z)\tilde\tau={1\over \tau^{\rm red}}T(u)\tau^{\rm
 red},
 \eea
 one can obtain after simple  calculations that the
 same relation (\ref{a15}) is valid.

 Now it is evident how to generalize this procedure to general multi-\mms
 using the formalism of scalar fields with ${\bf Z}_p$-twisted boundary
 conditions. It can be found in details in the references \cite{KMMMP,MP}.

 \subsection{Fermionic representations and forced hierarchies}
 \setcounter{equation}{0}
 \subsubsection{Determinant representations of $\tau$-functions}
 Beginning with this subsection we are going to spend more time discussing
 integrability in matrix models, which was practically out of our consideration
 before. In particular, now we are going to discuss the notion of
 $\tau$-function, main definitions and the determinant representations, which
we
 were talking about in the second subsection. We mainly follow the references
 \cite{DJKM,UT,KMMM}.

 We restrict ourselves to consideration of ordinary KP and Toda-lattice
 $\tau$-functions, which can be constructed as correlators in the theory of
 free fermions $\psi, \tilde\psi$ ($b,c$-system of spin $1/2$). The basic
 quantity is the ratio of fermionic correlators,

 \beq\label{tau}
 \tau _N(t,\bar t \mid G) \equiv \frac
 {\langle N \mid e^H G e^{\bar H} \mid N \rangle}
 {\langle N \mid  G  \mid N \rangle},
 \eeq
 in the theory of free 2-dimensional fermionic fields  $\psi (z)$,
 $\psi^{\ast} (z)$  with the action $\int
 \psi^{\ast} \bar \partial \psi $ (see the Appendix A1 for the definitions and
 notations). In the definition (\ref{tau})

 \beq\label{Ham}
 H = \sum_{k > 0} t_kJ_k; \ \ \ \bar H = \sum_{k>0} \bar t_k J_{-k},
 \eeq
 and the currents are defined to be

 \beq\label{cur}
 J(z) = \psi^{\ast} (z)\psi (z)\hbox{; }
 \eeq
 where

 \ba\label{psi}
 \psi (z) =
 \sum _{\bf Z}
 \psi _nz^n\ dz^{1/2}, \\
 \psi^{\ast} (z) =
 \sum _{\bf Z}
 \psi^{\ast} _nz^{-n-1}\ dz^{1/2}.
 \ea
 This expression for $\tau$-function is rather famous. Let us point out that
 this
 $\tau$-function is the $\tau$-function of the most general one-component
 hierarchy
 which is Toda lattice hierarchy. Putting all negative times to be zero as
 well as
 discrete index $n$, one obtains usual KP hierarchy. An element

 \beq\label{elGr}
  G =\
 :\exp \{ \sum_{m,n} {\cal G}_{mn}\psi^{\ast} _m\psi _n\}:
 \eeq
 is an element of the group $GL(\infty)$ realized in the infinite dimensional
 Grassmannian. The normal ordering should be understood here with respect to
 any $|k\rangle$ vacuum,\footnote {Indeed, it is rather crucial point and we
 will return to possible choices of the normal orderings.}
 where vacuum states are defined by conditions

 \beq\label{vac}
 \psi _m|k\rangle  = 0\ \ m < k\hbox{ , }  \psi^{\ast} _m|k\rangle  = 0\ \
 m \geq  k.
 \eeq
  Any particular solution of the hierarchy depends
 only on the choice of the element  $G$ (or, equivalently it can be uniquely
 described by the matrix  ${\cal G}_{km}$). Thus, we can reformulate the sense
 of constraint algebra in matrix models: it just fixes the element of the
 Grassmannian, i.e. one can reformulate constraint algebra in terms of
equations
 on the element of the Grassmannian. This approach was advocated in \cite{KS}.

 {}From commutation relations for the fermionic modes, one can conclude that
 any element in the form (\ref{elGr}) rotates the fermionic modes as follows

 \beq\label{A16}
 G\psi _kG^{-1} = \psi _jR_{jk}\hbox{ , }  G\psi ^\ast _kG^{-1} =
 \psi ^\ast _jR^{-1}_{kj} ,
 \eeq
 where the matrix  $R_{jk} $ can be expressed through  ${\cal G}_{jk}$ (see
 \cite{26}). We
 will see below that the general $\tau$-function (\ref{tau}) can be expressed
in
 the
 determinant form with explicit dependence of  $R_{jk}$ . In order to calculate
 $\tau $-function  we need some more notations. Using commutation relations
 for the fermionic modes, one can obtain the evolution of  $\psi (z)$ and
$\psi
 ^\ast (z)$ in
 times $\{t_k\}$, $\{\bar t_k\}$ in the form

 \beq\label{A17}
 \psi (z,t) \equiv  e^{H(t)}\psi (z)e^{-H(t)} = e^{\xi (t,z)}\psi (z)\hbox{  ,}
 \eeq
 \beq
 \label{A18}
 \psi ^\ast (z,t) \equiv  e^{H(t)}\psi ^\ast (z)e^{-H(t)} =
 e^{-\xi (t,z)}\psi ^\ast (z)\hbox{  ;}
 \eeq
 \beq
 \label{A19}
 \psi (z,\bar t) \equiv  e^{\bar H(\bar t)}\psi (z)e^{-\bar H(\bar t)} =
 e^{\xi (\bar t,z^{-1})}\psi (z)\hbox{ , }
 \eeq
 \beq
 \label{A20}
 \psi ^\ast (z,\bar t) \equiv  e^{\bar H(\bar t)}\psi ^\ast (z)e^{-\bar H(\bar
 t)} = e^{-\xi (\bar t,z^{-1})}\psi (z)\hbox{ ,}
 \eeq
 where

 \beq\label{A21}
 \xi (t,z) = \sum ^\infty _{k=1}t_kz^k\hbox{ .}
 \eeq

 Now let us define Shur polynomials by the formula:

 \beq\label{Shur}
 \exp\{\sum_{k>0}t_kx^k\} \equiv \sum _{k>0} P_k(t_k)x^k.
 \eeq
 Using this definition,
 from eqs. (\ref{A17})-(\ref{A20}) one can easily obtain the evolution of the
 fermionic modes:

 \beq\label{A22}
 \psi _k(t) \equiv e^{H(t)}\psi _ke^{-H(t)}
 =\sum ^{\infty}_{m=0} \psi _{k-m}P_m(t) ,
 \eeq
 \beq
 \label{A23}
 \psi ^{\ast }_k(t) \equiv e^{H(t)}\psi _k^{\ast }e^{H(t)} =
 \sum ^{\infty}_{m=0} \psi ^{\ast }_{k+m}P_m(-t) ;
 \eeq
 \beq\label{A24}
 \psi _k(\bar t) \equiv e^{\bar H(\bar t)}\psi
 _ke^{-\bar H(\bar t)} =\sum ^{\infty} _{m=0} \psi
 _{k+m}P_m(\bar t);
 \eeq
 \beq
 \label{A25}
 \psi _k^{\ast }(\bar t) \equiv e^{\bar H(\bar t)}\psi
 _k^{\ast }e^{-\bar H(\bar t)}=\sum ^{\infty} _{m=0} \psi
 _{k-m}^{\ast } P_m(-\bar t).
 \eeq
 It is useful to introduce the totally occupied state  $|-\infty \rangle $
 which satisfies the
 requirements

 \beq\label{A26}
 \psi ^\ast _i|-\infty \rangle  = 0\hbox{  , }  i \in  {\bf Z}\hbox{  .}
 \eeq
 Then any shifted vacuum can be generated from this state as follows:

 \beq\label{A27}
 |n\rangle  = \psi _{n-1}\psi _{n-2} ...|-\infty \rangle \hbox{  .}
 \eeq
 Note that the action of an any element  $G$  of the Clifford group (and, as
 consequence, the action of $e^{-\bar H(\bar t)} $) on  $|-\infty \rangle $  is
 very
 simple: $G|-\infty \rangle  \sim  |-\infty \rangle $, so using (\ref{A23}) and
 (\ref{A24}) one can obtain from eq.(\ref{tau}):

 \beq
 \new
 \begin{array}{c}
 \tau _n(t,\bar t) =
 \langle -\infty |...\psi ^\ast _{n-2}(-t)\psi ^\ast _{n-1}(-t)G\psi _{n-1}(-
 \bar t)\psi _{n-2}(-\bar t) ...|-\infty \rangle  \sim \\
 \sim \det [\langle -\infty |\psi ^\ast _i(-t)G\psi _j(-\bar t)G^{-1}|-\infty
 \rangle ]\left| _{i,j \leq  n-1}\right.\hbox{ .}
 \end{array}\label{A28}
 \eeq
 Using (\ref{A16}) it is easy to see that

 \beq\label{A29}
 G\psi _j(-\bar t)G^{-1} =
 \sum _{m,k} P_m(-\bar t)\psi _kR_{k,j+m}
 \eeq
 and the ``explicit" solution of the two-dimensional Toda lattice has the
 determinant representation:

 \beq\label{A30}
 \tau _n(t,\bar t) \sim \left. \det \ {H}_{i+n,j+n} (t,\bar t)
 \right| _{i,j<0}\hbox{ ,}
 \eeq
 where

 \beq\label{A31}
 {H}_{ij}(t,\bar t) = \sum _{k,m} R_{km}P_{k-i}(t)P_{m-j}(-\bar t)\hbox{  .}
 \eeq
 The ordinary solutions to KP hierarchy \cite{DJKM} correspond to the case when
 the
 whole evolution depends only of positive times $\{t_k\}$; negative times
 $\{\bar t_k\}$ serve as parameters which parameterize the family of points in
 Grassmannian and can be absorbed by re-definition of the matrix $R_{km}$. Then
 $\tau $-function of (modified) KP hierarchy has the form

 \beq\label{A32}
 \tau _n(t) = \left.\langle n|e^{H(t)}G(\bar t)|n\rangle \sim \det \left[\sum
 _k
 R_{k,j+n}(\bar t)P_{k-i-n}(t)\right]\right| _{i,j<0}\hbox{ ,}
 \eeq
 where  $G(\bar t) \equiv  Ge^{-\bar H(\bar t)}$ and

 \beq\label{A33}
 R_{kj}(y) \equiv  \sum  _m R_{km}P_{m-j}(-y)\hbox{  .}
 \eeq

 \subsubsection{Determinant representations for matrix model hierarchies}
 Now let us briefly discuss the formulas of the second subsection within the
 developed framework. Indeed, we already have the general determinant
 representation (\ref{A30}) and would have to reduce it to concrete
 subhierarchies. The life is, however, not so simple as all the determinants we
 know from matrix models are {\it finite} determinants, in contrast to
 (\ref{A30}). This is an essential problem. Indeed, this is due to the specific
 boundary condition like (\ref{forced}) imposed on matrix model
 $\tau$-functions. The hierarchies with this boundary condition are called
 forced hierarchy \cite{Kaup} and studied in details in \cite{KMMOZ,KMMM}. The
 main result of this study is that these hierarchies are corresponded by the
 singular elements of the Grassmannian, which are induced by the fermion
 operators (\ref{psi}), but with only positive (or only negative) modes taken
 into account. It implies the consideration in (\ref{elGr}) of {\it
 semi}-infinite (instead of infinite) matrix of the group $GL(\infty)$. In its
 turn, it leads to the {\it finite} (instead of semi-infinite) determinant
 (\ref{A30}).

 Indeed, there is the problem of continuing the boundary condition
 (\ref{forced}) to negative $n$. Say, the Toda chain integrable equations can
 not fix it (it can be trivially observed from Hirota equation - see
 (\ref{Hirota})). However, switching on the negative (Toda) times allows one to
 fix this continuation. And if one wishes to have the Toda chain as a proper
 reduction from Toda lattice hierarchy, it is necessary to put

 \beq\label{forcedcond}
 \tau_{-n}=0
 \eeq
 (the case of $\tau_{-n}=\tau_n$, for example, corresponds to CKP system - see
 \cite{KMMOZ}). To do this, one should multiply the element of the Grassmannian
 by the projector onto positive $n$:

 \beq\label{projector}
 P_+ = :\exp [\sum _{i<0}\psi _i\psi ^\ast _i]:,
 \eeq
 with the properties:

 \ba\label{prpro}
 P_+|n\rangle  = \theta (n)|n\rangle ,\\
 P_+\psi ^\ast _{-k} = \psi _{-k}P_+ = 0\hbox{  , }  k > 0\ ;\\
 \left[ P_+,\psi _k\right] = [P_+,\psi ^{\ast} _k] = 0\hbox{  , } k \geq  0\ .
 \ea
 The point, however, is that after this is done, the defining properties of the
 original (non-forced) system are the same as those of the forced one under the
 condition (\ref{forcedcond}) (i.e. after throughing away the half of fermionic
 modes and inserting the projector). Therefore, we discuss now the Toda lattice
 hierarchy and its Toda chain reduction with no restricting ourselves to the
 forced case.

 Let us note that the entries (\ref{A31}) of the determinant (\ref{A30}) are
not
 arbitrary ones, but are conditioned to satisfy the properties:

 \bea\label{h1}
 \partial H_{ij}/\partial t_p = H_{i,j-p}\hbox{, }   j>p>0,\\
 \label{h2}
 \partial H_{ij}/\partial \bar t_{p} = - H_{i-p,j}\hbox{, }   i>p>0,
 \eea
 which are the automatic consequences of the corresponding property of Shur
 polynomials:

 \beq\label{h3}
 \partial P_k/\partial t_p = P_{k-p}
 \eeq
 following immediately from their definition. From the properties (\ref{h1})
and
 (\ref{h2}) one obtains directly our conditions (\ref{cond1}) and (\ref{cond3})
 and their analogs for the multi-matrix case (\ref{detrepMMM}) (in the latter
 case, one should change the sign in the last potential in (\ref{MMM}), which
is
 fixed by the sign in the Hamiltonians in $\bar t_k$). Now let us discuss the
 condition (\ref{cond2}), i.e. the reduction of general Toda lattice hierarchy
 to Toda chain.  It can be easily
 formulated as condition to the element $G$ of the Grassmannian:

 \beq\label{h4}
 [J_k + \bar J_k,G] = 0,
 \eeq
 which is equivalent to constraint

 \beq\label{h5}
 [\Lambda  + \Lambda ^{-1},R] = 0\hbox{  .},
 \eeq
 where $\Lambda $ is shift matrix  $\Lambda _{ij}\equiv \delta _{i,j-1}$. In
 this case,

 \beq\label{h6}
 Ge^{-\bar t_k\bar J_k} = e^{-\bar t_k\bar J_k} e^{-\bar t_kJ_k} G\ e^{\bar
 t_kJ_k}
 \eeq
 and $\tau $-function depends (up to the trivial factor) only on times  $\{t_k
-
 {\bar t_k}\}$:

 \beq\label{h7}
 \tau _n(t,{\bar t}) = e^{\sum kt_k\bar t_k} \langle n|e^{H(t-\bar
t)}G|n\rangle
 \hbox{  .}
 \eeq
 The reduction (\ref{h5}) has an important
 solution\footnote{Generally the solutions
 $R_{nk}=R_{n+k}$ and  $R_{nk}=R_{n-k}$ are
 different, but for forced hierarchy, when $\tau $-function is the determinant
 of finite matrix, these two solutions are equivalent due to possibility
 to reflect
 matrix with respect to vertical axis without changing the determinant.}

 \beq\label{h8}
 R_{km} = R_{k+m}\hbox{ .}
 \eeq
 In this case the matrix ${\hat H}_{ij}$ defined by eq.(\ref{A31}) evidently
 satisfies
 the
 relations ${H}_{ij} = {H}_{i+j}$ and

 \beq\label{h9}
 (\partial _{t_k}+ \partial _{\bar t_k}){H}_{i+j} = 0
 \hbox{     for any }\ \ k<n-i
 \ ,\ k<n-j
 \eeq
 due to the properties (\ref{h1}) and (\ref{h2}).
 The property (\ref{h9}) certainly does not imply that the corresponding
 $\tau $-function depends only on difference of times because of restriction of
 values of $k$, but it restores correct dependence of times with taking into
 account of exponential in (\ref{h7}).

 Thus, we have proved that, for the Toda chain hierarchy, the matrix $H_{ij}$
 satisfies the condition

 \beq\label{h10}
 [H\hbox{ , } \Lambda +\Lambda ^{-1}]=0,
 \eeq
 which leads to $\tau $-function of Toda chain hierarchy (properly rescaled by
 exponential of bilinear form of times) which depends only on the difference of
 positive and negative times  $t_p-\bar t_{p}$,
 but not on their sum (one can
 consider this as defining property of Toda chain hierarchy). Let us remark
that
 only one possible solution to constraint (\ref{h10}) is matrix
 $H_{i,j}={H}_{i+j}$, but it
 should not be generally independent of the difference of times  $t_p-\bar
 t_{p}$. Certainly, one can through out negative times as the final answer for
 complete objects like $\tau $-function should be really independent of this
 difference.

 Anyway, we obtained for the $\tau$-function of Toda chain hierarchy the
 determinant representation with necessary conditions
 (\ref{cond1})-(\ref{cond3}), i.e. the matrix $H_{ij}$ just corresponds to
 moment matrix $C_{ij}$ and we finally reproduce the result (\ref{detrepTC}).

 \subsubsection{$\ast$ Integrable structure of CMMM - multi-component
hierarchy}
 Now we are going to say some words on integrable structure which arises in
 CMMM. In fact, we already wrote down the determinant representation
 (\ref{CMMMdetrep}) for CMMM. Now there are two things which can be
established.
 The first one is to obtain the integrable equations. It can be done
immediately
 from the equation (\ref{CMMMdetrep}) and leads to clear generalization of the
 Hirota equation (\ref{Hirota}):

 \beq\label{HirotaCMMM}
 {1\over \tau _{N,M}}{\partial ^2\over \partial t_1\partial \bar t_1} \tau
 _{N,M}) - \left( {\partial \tau _{N,M} \over \partial t_1}\right)  \left(
 {\partial \tau _{N,M}\over \partial \bar t_1}\right)=
 {\tau _{N+1,M-1}
 \tau _{N-1,M+1}\over \tau ^2_{N,M}}.
 \eeq
 The second point is just the properties of the $\tau$-function corresponding
to
 CMMM within the framework described above. The statement is that the partition
 function of CMMM is a $\tau $-function of a multi-component
($(p+1)$-component)
 Kadomtsev-Petviashvili hierarchy which obeys the constraint

 \beq\label{34}
 \sum ^{p+1}_{k=1}\partial /\partial t^{(k)}_n\tau ^{(p+1)}(\{t\}) = 0,  \ \ \
    n =  1,2,\ldots\  ,
 \eeq
 where  $p$  is the number of matrices in CMMM. Let us demonstrate how it works
 in our usual simplest example of $p=1$. It should be 2-component hierarchy.

 The $\tau $-function of 2-component KP
 hierarchy is by definition the correlator

 \beq\label{3.2.1}
 \tau ^{(2)}_{N,M}(t,\bar t) = \langle N,M|e^{H(t,\bar t)}G|N+M,0\rangle
 \eeq
 where

 \beq\label{3.2.2}
 H(t,\bar t) = \sum _{k>0}(t_kJ^{(1)}_k + \bar t_kJ^{(2)}_k)
 \eeq
 \beq\label{3.2.3}
 J^{(i)}(z) = \sum    J^{(i)}_kz^{-k-1} = {:}\psi
 ^{(i)}(z) \psi ^{(i)\ast }(z){:}
 \eeq
 \beq\label{3.2.4}
 \psi ^{(i)}(z) \psi ^{(j)\ast }(z') = {\delta _{ij}\over z - z'} + \ldots\ \ .
 \eeq
 All these formulas are trivially generalized to $(p+1)$-component case by
 introducing $p+1$ fermions.

 Now we are going to demonstrate that (\ref{z2cor}) is equivalent to
 (\ref{3.2.1})
  for certain
 $G$  for which (\ref{3.2.1}) depends only on the differences  $t_k-\bar t_k$.
 To do
 this we have to make use of the free-fermion representation of  $SL(2)_{k=1}$
 Kac-Moody algebra:

 \bea\label{3.2.5}
 &J_0 = {1\over 2}(\psi ^{(1)}\psi ^{(1)\ast } - \psi ^{(2)}\psi ^{(2)\ast }) =
 {1\over 2}(J^{(1)} - J^{(2)})\nn\\
 &J_+ = \psi ^{(2)}\psi ^{(1)\ast } \quad J_- = \psi ^{(1)}\psi ^{(2)\ast }
 \eea
 Now let us take $G$ to be the following exponent of a quadratic form (indeed,
 we know that this is the correct fermionic form for exponential of screening
 operators)

 \beq\label{3.2.6}
 G \equiv  {:}\exp  \left( \int   \psi ^{(2)}\psi ^{(1)\ast }\right){:}
 \eeq
 Now we bosonize the fermions
 \bea\label{3.2.8}
 \psi ^{(i)\ast } = e^{\phi _i},  \ \ \  \psi ^{(i)} = e^{- \phi _i}
 \nn\\
 J^{(1)} = \partial \phi _1, \ \ \     J^{(2)} = \partial \phi _2
 \eea
 and compute the correlator

 $$
 \tau ^{(2)}_N(t,\bar t) \equiv  \tau ^{(2)}_{N,-N}(t,\bar t) =
 $$
 $$
 = {1\over N!}
 \langle N,-N|\exp \left( \sum _{k>0}(t_kJ^{(1)}_k +
 \bar t_kJ^{(2)}_k)\right) \left( \int
 {:}\psi ^{(2)}\psi ^{(1)\ast }{:}\right) ^N|0\rangle  =
 $$
 $$
 = {1\over N!} \langle N,-N|\exp \left(
 \oint
 [V_1(z)J^{(1)}(z) + V_2(z)J^{(2)}(z)]\right) \left( \int
 :\exp (\phi _1-\phi _2):\right) ^N|0\rangle,
 $$
 where $V_{(1,2)} \equiv \sum_k (t_k,\bar t_k)z^k$.
 Introducing the linear combinations  $\sqrt{2}\phi  = \phi _1 - \phi _2$,
 $\sqrt{2}\tilde \phi  = \phi _1 + \phi _2$ we finally get

 \ba\label{3.2.9}
 \tau ^{(2)}_N(t,\bar t) = {1\over N!} \langle \exp \left( {1\over \sqrt{2}}
 \oint[V_1(z)+V_2(z)]\partial \tilde \phi (z)\right) \rangle  \times\\
 \times  \langle N|\exp \left( {1\over \sqrt{2}}
 \oint [V_1(z)-V_2(z)]\partial \phi (z)\right) \left( \int
 :\exp \sqrt{2}\phi :\right) ^N|0\rangle  = \tau ^{(2)}_N(t-\bar t)
 \ea
 since the first correlator is in fact independent of $t$ and $\bar t$. Thus,
we
 proved that the $\tau $-function (\ref{3.2.1})
 indeed depends only on the {\it difference} of two sets
 of times  $\{t_k-\bar t_k\}$. So, we obtained here a particular case of the
 2-component KP hierarchy (\ref{3.2.1}) and

 (i)  requiring the elements of Grassmannian to be of the form (\ref{3.2.6})
 we actually performed a reduction to the 1-component case\footnote{
 Note that the idea to preserve both indices in (\ref{3.2.1}) leads
 immediately to additional insertions either of  $\psi^{(1)\ast}$ or
 $\psi ^{(2)}$ to the right vacuum $|0\rangle $, so that it is no longer
 annihilated at least by  the $T_{-1}$ Virasoro generator, or in other words
 this ruins the string equation. Thus only the particular reduction
 (\ref{3.2.6}) seems to be
 consistent with string equation. This choice of indices just corresponds to
 that considered originally in \cite{DJKM}.};

 (ii)  we proved in (\ref{3.2.9}) that this is an AKNS-type reduction for
 the $\tau $-function (\ref{3.2.8}) \cite{UT,Newell}.

 The above simple example already contains all the basic features of at least
 all the  $A_p$ cases. Indeed, the reduction (\ref{3.2.9})
 is nothing but  $SL(2)$-reduction
 of a generic  $GL(2)$  situation. In other words, the diagonal  $U(1)$
 $GL(2)$-current  $\tilde J = {1\over 2}(J^{(1)} + J^{(2)}) =
 {1\over \sqrt{2}}\partial \tilde \phi $  decouples. This is an invariant
 statement which can be easily generalized to higher  $p$  cases.
 Indeed, the screening operators for the $p$-matrix case are the integrals of
 $SL(p+1)_1$ Kac-Moody currents (not  $GL(p+1)$ ones)
 and, thus, the $\tau $-function (\ref{zcmm})-(\ref{screen}) does not depend on
 $\{\sum ^{p+1}_{i=1}x^{(i)}_k\}$, i.e. we obtain the constraint (\ref{34}).

 As to the Toda-like representation of CMM, in the simplest $SL(2)$-case
 the result should be equivalent to the Toda chain hierarchy. In
 the fermionic language this connection is established by the following
 substitution in the element of the Grassmannian

 \beq\label{3.2.20}
 \psi ^{(1)}(z) \rightarrow  \psi (z), \ \ \
 \psi ^{(2)}(z) \rightarrow  \psi ({1\over z})
 \eeq
 and the same for $\psi ^\ast$'s. This is a reflection of the fact that Toda
 system is described by the two marked points (say, 0 and $\infty$) and
 corresponds to two glued discs, so it can be also described by two different
 fermions. This might lead to a general phenomenon, when any multi-component
 solution to CMM is actually related to (some reduction) of a multi-component
 Toda lattice.

\section{GKM approach to matrix models}
\subsection{What are the matrix models in the double scaling limit}
\setcounter{equation}{0}
In this section we are going to discuss the models which arise {\it after}
the continuum (double scaling) limit is taken. Indeed, now we are merely
formulate a set of conditions, which uniquely defines the corresponding
partition function.

These conditions are of two kinds. The first one describes the integrable
properties. That is, the square root of the partition function of the
Hermitean $K$-matrix model in the continuum limit is a $\tau$-function of
the $K$-reduced KP hierarchy. Say, one-matrix model corresponds to KdV
hierarchy, two-matrix model -- to Boussinesq hierarchy and etc.

The second condition imposed on the partition function is so-called string
equation. The role of this equation is to fix the $\tau$-function which
corresponds to the matrix model.
Put it differently, this equation gives the point of the Grassmannian which
corresponds to the matrix model. The equation is

\beq
{\partial\over\partial x}{\cal L}_{-1}^{\{K\}}\tau=0,
\eeq
where

\beq\label{LFKN}
{\cal L}_{-1}^{\{K\}}={\f K}\sum_{n>K} nT_n{\partial\over\partial T_{n-K}}+
{\f 2K}\sum_{{a+b=K}\atop{a,b>0}} aT_abT_b + {\partial\over\partial T_1}.
\eeq
Indeed, hereafter we call string equation a bit more general equation

\beq\label{streq}
{\cal L}_{-1}^{\{K\}}\tau=0,
\eeq
which is also correct.

It can be demonstrated \cite{FKN2} that above mentioned conditions imply
that the $\tau$-function satisfies a whole set of constraints. More concretely,
$K$-reduced $\tau$-function satisfies the set of W-algebra constraints:

\beq
W_n^{(i)}\tau^{\{K\}}=0,\ \ i=2,...,K;\ \ n\ge -i+1.
\label{W}
\eeq
$W$-operators here are the standard generators of Fateev-Lukyanov-Zamolodchikov
$W$-algebra \cite{FL} expressed in the manifest terms of times, i.e. with
creation and annihilation operators being $T_n$ and $\partial/\partial T_n$
respectively and acting on the spce of function of times \cite{FKN}.

There is also inverse conjecture \cite{FKN} which asserts that the
constraints (\ref{W}) have a unique solution, i.e. they select out the
$\tau$-function of $K$-reduced KP hierarchy, which satisfies the string
equation
(\ref{streq}). This conjecture is still not proved, though there are strong
arguments in favor of its correctness \cite{GKMU}.

Let us note now that it is not so simple to prove the properties described
above, and, in part, they are merely an invariant definition of the double
sclaing continuum limit. Some general argument have been used to derive these
properties \cite{DVV,FKN}, but there are still only two rigid calculations.

The first one describes the transition from discrete to continuum $W$- and
Virasoro constraints and has been done for $K=2$ case of the standard
matrix models \cite{MMMM} and for the
 general case of the conformal multi-matrix
models \cite{KMMMP,MP}.
Another calculation makes use of the manifest matrix integral representation
of the partition function \cite{KMMM}, which will be described in the next
subsections.

\subsection{Generalized Kontsevich Model (GKM)}
\setcounter{equation}{0}
Thus far we have described the set of constraints which uniquely defines the
partition function of matrix models in the double scaling limit. Now we are
going to present the manifest solution of these constraints. Indeed, for
the simplest case of one-matrix model it was derived from Witten topological
theory (see section 3) by Kontsevich \cite{Kon2}.
We propose proper generalization of
his result, demonstrating that our solution satisfies all necessary
constraints (in contrast to the approach \cite{Kon2}, where there has been
manifestly  calculated the
integral (\ref{topcor})).

For the lack of space, we will briefly describe only the structures and
appealing properties of the partition function,
which
we call {\it Generalized Kontsevich's Model} (GKM) \cite{KMMMZ}.
Unfortunately, we can not explain here the deep connections of GKM with
the theory of integrable systems and develop possibile generalizations,
valuable in different applications. In particular, the connection with
topological $N=2$ Landau-Ginzburg theories is out of the scope of the
present review. We refer to the conclusion for a short review of the
existing literature.

The partition function of the GKM is
defined by the following integral over $N \times N$ Hermitean matrix:

\beq\label{GKM}
Z^{\{{\cal V}\}}_N[M] \equiv  \frac{\int e^{U(M,Y)}dY}
{\int e^{-U_2(M,Y)}dY} \hbox{ , }
\eeq
where

\beq
U(M,Y) = Tr[{\cal V}(M+Y) - {\cal V}(M) - {\cal V}\ '(M)Y]
\eeq
and

\beq
U_2(M,Y) =   \lim_{\epsilon \rightarrow 0}
{1\over \epsilon ^2} U(M,\epsilon Y) \hbox{ , }
\eeq
is an $Y^2$-term in $U$. $M$ is also a Hermitean $N \times N$ matrix with
eigenvalues $\{\mu _i\}$,
${\cal V}(\mu)$ is arbitrary analytic function.

All crucial properties of GKM follows from its definition. That is,
its partition function may be
considered as a functional of two different variables:  potential
${\cal V}({\mu})$ and the infinite-dimensional Hermitean matrix $M$ with
eigenvalues $\{{\mu}_i\}$. Partition function $Z^{\{{\cal V}\}}_N$ is an
$N$-independent KP $\tau $-function, considered as a function of time-variables
$T_n= {1\over n}TrM^{-n}$ and the point of Grassmannian is specified by the
choice of potential. The $N$-dependence enters only through the argument  $M:$
we return to finite-dimensional matrices if only $N$ eigenvalues of $M$ are
finite. In this sense the ``continuum" limit of  $N \rightarrow  \infty $  is
smooth.

The GKM is associated with a subset of Grassmannian, specified by additional
${\cal L}_{-1}$-constraint. For particularly adjusted potentials
${\cal V}({\mu}) = const\cdot {\mu}^{K+1}$, the corresponding points in
Grassmannian lies in the subvarieties, associated with $K$-reductions of
KP-hierarchy,  $Z^{\{{\cal V}\}}$ becomes independent of all the time-variables
$T_{Kn}$, and the  ${\cal L}_{-1}$-constraint implies the whole tower of
$W_K$-algebra constraints on the reduced $\tau $-function. These properties are
exactly the same as suggested for double scaling limit of the $K-1$-matrix
model, and in fact there is an identification

\beq
Z_{\infty}^{\{K\}}=\sqrt{\Gamma_{ds}^{\{K-1\}}}.
\eeq

All this means that GKM provides an interpolation between double-scaling
continuum limits of all multimatrix models and thus between all string models
with $c\leq 1$. Moreover, this is a reasonable interpolation, because both
integrable and ``string-equation" structures are preserved. This is why we
advertise GKM as a plausible (on-shell) prototype of a unified theory of 2d
gravity.

In the next subsections we briefly comment all properties described here.

\subsection{Main integrable properties of GKM}
\setcounter{equation}{0}
\subsubsection{Integrable structure}
After the shift of variables $X=Y+M$ and
integration over angular components of $X$,
$Z^{\{{\cal V}\}}_N[M]$ acquires the form of

\beq\label{hn4}
Z^{\{{\cal V}\}}_N[M] = {[\det
\tilde \Phi _i(\mu_j)]\over \Delta (M)}\hbox{,}
\eeq
where  $\Delta (M) = \prod _{i<j}(\mu_i-\mu_j)$  is the Van-der-Monde
determinant, and functions

\beq\label{hn5}
\tilde \Phi _i(\mu) = [{\cal V}\ ''(\mu)]^{1/2}
e^{{\cal V}(\mu)-\mu{\cal V}\ '(\mu)}\int
e^{-{\cal V}(x)+x{\cal V}'(\mu)} x^i dx
\eeq
The only assumption necessary for the derivation of (4) from (1) is the
possibility to represent the potential  ${\cal V}(\mu)$  as a formal series
in positive {\it integer} powers of $\mu$.

Formula (\ref{hn4}) with {\it arbitrary} entries  $\phi _i(\mu)$ having the
asymptotics

\beq\label{asymp}
\phi_i{\mu}\sim\mu^{i-1}(1+{\cal O}(\mu))
\eeq
is characteristic
for generic KP $\tau $-function  $\tau ^G(T_n)$  in Miwa's coordinates

\beq
T_n = {1\over n} TrM^{-n}\hbox{ , } n \ge 1
\eeq
and the point $G$ of Grassmannian is defined by potential $\cal V$
through the set of basis vectors
$\{\phi _i(\mu)\}$.
The most immediately this set is connected with the Baker-Akhiezer (BA)
function

\beq\label{BA}
\Psi_{\pm}(z|T_k)=e^{\sum
T_kz^k}{\tau(T_n\pm{z^{-n}\over n})\over \tau (T_n)}
\eeq
through the relation:

\beq
\Psi_+(\mu|T_k)\left|_{{T_1=x}\atop{T_k=0,\ k\ne 1}}=\sum_l {x^l\over l!}
\phi_l(\mu).\right.
\eeq

Thus, we obtain

\beq
Z^{\{{\cal V}\}}[M] = \tau ^{\{{\cal V}\}}(T_n).
\eeq
The case of finite $N$ in this formalism is distinguished by the condition that
only $N$ of the parameters $\{\mu_i\}$ are finite. In order to take the
limit $N \rightarrow  \infty $  in the GKM (\ref{GKM})
it is enough to bring all the
$\mu_i's$ from infinity. In this sense this a smooth limit, in contrast
to the singular conventional double-scaling limit, which one needs to take in
ordinary (multi)matrix models.

Let us check that this limit is really smooth. To do this, one should change
the number of Miwa variables $N+1\to N$ bringing $\mu_{N+1}$ to $\infty$ in
the determinant (\ref{hn4}). Using the asymptotics (\ref{asymp}), it is
trivially to prove that it results in $N\times N$ determinant of the same type.
This means that $N$ enters only as the number of Miwa variables, but not as an
explicit parameter.

\subsubsection{Reductions}
The integral  ${\cal F}^{\{{\cal V}\}}[\Lambda ]$,  $\Lambda
\equiv  {\cal V}\ '(M)$, in the numerator of (\ref{GKM})
satisfies the Ward identity

\beq\label{13}
Tr\left\lbrace \epsilon (\Lambda )\left[ {\cal V}\ '({\partial \over %
\partial \Lambda _{tr}}) - \Lambda \right] \right\rbrace
{\cal F}^{\{{\cal V}\}}_N =  0
\eeq
(as result of invariance under any shift of integration variable $X
\rightarrow  X + \epsilon (M))$. Let us denote the integral in (\ref{hn5})
through $F_i({\cal V}'(\mu))$. If potential ${\cal V}({\mu})$ is restricted
to be a polynomial of degree $K+1$, this identity implies that the functions
$F_i(\lambda)$ obey additional relations:

\beq
F_{m+Kn}(\lambda) =
\lambda ^n\cdot F_m(\lambda) +
\sum ^{m+Kn-1}_{i=1}s_iF_i(\lambda) \hbox{ . }
\eeq
Since the sum at the $r.h.s$. does not contribute to determinant (\ref{hn4}),
we can
say
that all the functions  $F_n$ are expressed through the first  $K$  functions
$F_1...F_K$ by multiplication by powers of  $\lambda = {\cal V}\ '({\mu})$.
Such
situation (when the basis vectors  $\phi _i$, defining the point of
Grassmannian are proportional to the first  $K$ ones) corresponds to reduction
of KP-hierarchy. This reduction depends on the form of  ${\cal V}\ '({\mu})$
and in the case of ${\cal V}({\mu})  = {\cal V}_K({\mu}) =
const\cdot {\mu}^{K+1}$ coincides with the well-known  $K$-reduction of the
KP-hierarchy (KdV as  $K=2$, Boussinesq as  $K=3$ etc.). Thus in such cases
partition function of GKM becomes  $\tau ^{\{K\}}$-function of the
corresponding hierarchy. Generic $\tau ^{\{K\}}$ possesses an important
property: it is
almost independent of all time-variables  $T_{nK}$. To be exact,

\beq
\partial \ \log\tau ^{\{K\}}/\partial T_{nK} = a_n = const.
\eeq
In variance with generic $\tau^{\{K\}}$, the partition function $Z^{\{K\}}$ of
GKM obeys this condition with all $a_n=0$ (see proof in \cite{KMMMZ}).

\subsection{String equation}
\subsubsection{${\cal L}_{-1}$-constraint}
The set of function
$\{\tilde \Phi _i({\mu})\}$ in (4) is, however, not arbitrary. They are all
expressed through a single function --- potential ${\cal V}({\mu}),- and$ are
in fact recurrently related:

\beq
F_i(\lambda ) = (\partial /\partial \lambda )^{i-1}F_1(\lambda )\hbox{.}
\eeq
This relation is enough to prove that

\beq
{\partial \over \partial T_1}\log \ Z^{\{{\cal V}\}}_N
 = - Tr\ M + Tr {\partial \over \partial \Lambda _{tr}}
\log \ \det \ F_i(\lambda _j)
\eeq
whenever potential ${\cal V}({\mu})$ grows faster than $\mu$ as $\mu
\rightarrow  \infty .$

Thus, $Z ^{\{{\cal V}\}}$ satisfies a simple identity:

\beq\label{9}
{1\over Z^{\{{\cal V}\}}} {\cal L}^{\{{\cal V}\}}_{-1}Z^{\{{\cal V}\}}_N =
{\partial \over \partial T_1}\log \ Z^{\{{\cal V}\}}_N + TrM -
Tr{\partial \over \partial \Lambda _{tr}}\log \ \det \ F_i(\lambda _j) = 0
\eeq
where operator ${\cal L}^{\{{\cal V}\}}_{-1}$ is defined to be

\bea\label{L}
{\cal L}^{\{{\cal V}\}}_{-1}= \sum _{n\geq 1}Tr
[{1\over {\cal V}\ ''(M)M^{n+1}}] {\partial \over \partial T_n} +
\nn \\
+ {1\over 2} \sum _{i,j}{1\over {\cal V}\ ''({\mu}_i){\cal V}\ ''({\mu}_j)}
{{\cal V}\ ''({\mu}_i)-{\cal V}\ ''({\mu}_j)\over {\mu}_i -  {\mu}_j} +
{\partial \over \partial T_1}
\eea
(the items with $i=j$ are included into the sum).
{}From eqs.(\ref{9}),(\ref{L}) it follows that
partition function of GKM satisfies the constraint

\beq\label{12}
{\cal L}^{\{{\cal V}\}}_{-1}\tau ^{\{{\cal V}\}} = 0.
\eeq

The reason why the operator (\ref{L}) is denoted by ${\cal L}^{\{{\cal
V}\}}_{-1}$ is clear from the following argument.
If ${\cal V} = {\cal V}_K$, the generic expression (\ref{L}) for the
${\cal L}_{-1}$-operator turns into

\beq\label{16}
{\cal L}^{\{K\}}_{-1} = {1\over K} \sum _{n>K} nT_n\partial /\partial T_{n-K} +
{1\over 2K }\sum_{{a+b=K}\atop {a,b>0}}
aT_abT_b + \partial /\partial T_1
\eeq
The last item at the $r.h.s$. may be eliminated by the shift of time-variables:

\beq
T_n \rightarrow  \hat T^{\{K\}}_n = T_n + {K\over n} \delta _{n,K+1}.
\eeq
Only expressed in terms of  these $\hat T$'s the constraint (\ref{L})
acquires the
form of

\ba
{\cal L}^{\{K\}}_{-1}\tau ^{\{K\}} =\\= \left\lbrace {1\over K} \sum _{{n>K}
\atop
{n \ne 0 mod K}}
n\hat T_n\partial /\partial \hat T_{n-K} + {1\over 2K}
\sum _{{a+b=K}\atop {a,b>0}}a\hat T_ab\hat T_b\right\rbrace  \tau ^{\{K\}}
=0 \hbox{ . }
\ea
coinciding with expression (\ref{streq}).

\subsubsection{Universal string equation}
Generalization of the string equation to the case of
arbitrary potential

\beq
{\partial \over \partial T_1}
{{\cal L}^{\{\cal V\}}_{-1}\tau ^{\{\cal V\}}\over \tau ^{\{\cal V\}}} = 0
\hbox{ . }
\eeq
may be transformed to the following form

\beq
\sum_{n\ge -1} {\cal T}_n \frac{\partial ^2 \log \tau}{\partial T_1
\partial T_n} = u \hbox{ , }
\eeq
where

\bea
{\cal T}_n \equiv Tr {1 \over V''(M)}{1 \over M^{n+1}} \hbox{ , } \\
u \equiv \frac{\partial ^2 \log \tau }{(\partial T_1 )^2}\hbox{ , }
\frac{\partial \log \tau }{\partial T_0} \equiv 0 \hbox{ , }
\frac{\partial \log \tau }{\partial T_{-1}} \equiv T_1 \hbox{ . } \nn
\eea
Using BA function (\ref{BA}), one can rewrite string equation (\ref{12})
in the form of bilinear relation

\beq
\sum_i \frac{\Psi _+ (\mu_i) \Psi _-(\mu _i)}{\mu _i} = u \hbox{ . }
\eeq

\subsubsection{$\cal W$-constraints}
According to subsection 5.1 the
constraint

\beq\label{24}
{\cal L}^{\{K\}}_{-1}\tau ^{\{K\}} = 0
\eeq
implies the entire tower of $\cal W$-constraints

\beq\label{25}
{\cal W}^{(k)}_{Kn} Z^{\{K\}} = 0\hbox{, }   k = 2,3,...,K\hbox{; }  n\geq  1-k
\eeq
imposed on  $\tau ^{\{K\}}$. Here ${\cal W}^{(p)}_{Kn}$ is the $n-th$ harmonics
of the  $p-th$ generator of Zamolodchikov's  $W_K$-algebra (the proper notation
would be  ${\cal W}^{(p)\{K\}}_n$, but it is a bit too complicated). There is a
Virasoro Lie sub-algebra, generated by ${\cal W}^{(2)}_{Kn} =
{\cal L}^{\{K\}}_n$, and the particular ${\cal L}^{\{K\}}_{-1}$ is just the
operator (\ref{16}). This is the true origin of our notation
${\cal L}^{\{{\cal V}\}}_{-1}$ in the generic situation (where the entire
Virasoro subalgebra of $W_\infty $ was not explicitly specified).

Besides being a corollary of (\ref{24}), the constraints (\ref{25})
can be directly deduced
from the Ward identity (\ref{13}). For the case of  $K=2$  (which is original
Kontsevich's model \cite{Kon2}) this derivation was given in \cite{singer}
 (see also\cite{GN,Wit,MakSem}
for
alternative proofs). Unfortunately, for $K\geq 3$ the direct corollary of
(\ref{13}) is not just (\ref{25}),
but peculiar linear combinations of these constraints,
$e.g$. for  $K=3$  they look like

\ba\label{26}
{\cal W}^{(3)}_{3n}Z^{\{3\}}_\infty  = 0\hbox{, }     n\geq -2;
 \\
\left\lbrace \sum _{k\geq 1}(3k-1)\hat T_{3k-1}{\cal W}^{(2)}_{3k+3n}
+\sum _{a+b=3n} {\partial \over \partial T_{3a+2}}
{\cal W}^{(2)}_{3b-3}\right\rbrace  Z^{\{3\}}_\infty  = 0\hbox{, }\\
a,b\geq 0\hbox{, }  n\geq -2;\\
\left\lbrace \sum _{k\geq 1+\delta _{n+3,0}}(3k-2)\hat T_{3k-2}%
{\cal W}^{(2)}_{3k+3n} +\sum _{a+b=3n}
{\partial \over \partial T_{3a+1}} {\cal W}^{(2)}_{3b-3}\right\rbrace
Z^{\{3\}}_\infty  = 0\hbox{, } \\ a,b\geq 0\hbox{, }  n\geq -3.
\ea
For identification of (\ref{26}) with (\ref{25}) one can argue
that both sets of
constraints possess unique, and thus coinciding, solutions.
Really, only the case of $K=3$ has been investigated in this case in the paper
\cite{Mik}.

\section{Conclusion}
\setcounter{equation}{0}
In the present review we have discussed in some details only Liouville
(physical) gravity
and discrete matrix models. We almost ignored topological theories, and
described
GKM approach very briefly, though they deserve separate reviews.

Indeed, this is the
GKM approach which allows one to connect different branches and
viewpoints. On one hand, for the particular case of the cubic potential ($K=2$)
GKM integral can be immediately derived from the Witten topological theory
\cite{Kon}.
On the other hand, it is possible to establish connection between GKM and
$N=2$ Landau-Ginzburg topological gravity \cite{LGM,LosPol}. As
topological theories of both these types are appropriate models of $2d$
gravity,
the connections of these with GKM are of great importance.

Let us note that there are many generalizations of the original GKM integral
(\ref{GKM}). Say, one can include "negative" and "zero" time variables to
obtainToda lattice $\tau$-functions \cite{KMMM}. It allows to incorporate
in GKM treatment Hermitean one-matrix model and to take the
double scaling limit immediately in the partition function \cite{KMMM,ACKM}.

Another possible generalization is to consider polynomials of ${\f X}$ as the
GKM potential. This case has much to do with unitary matrix integrals
\cite{GKMU}.

An absolutely different issue is the existence of different phases in GKM. The
new one can be defined by an other Miwa transformation of times:

\beq
\tilde T_k\equiv {\f k} \hbox{\rm Tr} M^k.
\eeq
In fact, it rather determines a different aymptotics of GKM
integral than a new phase. It acquires the sense of phase only when
GKM integral is applied to the description of QCD \cite{GW}.

A separate issue is the connection between GKM integral and
character expansion of $\tau$-functions
\cite{GKM2}. This point is rather important in
applications both to unitary matrix models and to $2d$ QCD
\cite{GKM2,GKMU}.\footnote{In particular,
the applications to the $2d$ QCD case are connected with the
fact that one- and two-point correlation functions in this theory
are $\tau$-functions
of KP and Toda lattice hierarchies respectively.}

Another limitation of this review is that
we restricted ourselves to Hermitian {\it only} matrix models and {\it only}
to one-cut solution in the continuum limit. In fact, the properties, say, of
unitary matrix model \cite{BG,GW,PS,GMMMO}
is understood well too. Its integrable
properties are described at discrete level \cite{KM} as well as in the
continuum limit \cite{BS,HMNP,GKM2}. Proper
Virasoro algebra was also investigated, again both at the discrete
\cite{BMS,MM} and continuum
\cite{GN} levels. Different reductions were discussed \cite{BMS} and
GKM framework proposed \cite{GN,GKMU}.

Nevertheless, these models have no clear physical implications for
$2d$ gravity, but
rather in $2d$ QCD. This model is also not included into physically essential
general framework, which would be multi-component hierarchies. These have
(presumably) to include at equal footing DKP hierarchy and some other
reductions having to do with different $2d$ gravity theories, including those
with $c=1$. These models should be associated with non-trivial critical points
like multi-cut solutions \cite{DC}.

Thus, unitary model, on one hand, requires a separate review and, on the other
hand, is out of our main line. This is why we avoided to discuss them here.

This discussion might explain that it is impossible to put all the
material on the small room of this review.
The above-mentioned gaps will be filled in the
second part of the review, devoted exclusively to matrix models in external
fields, which have much to do both with $2d$ gravity (see section 5 above) and
$2d$ gauge theories.

\section*{Acknowledgements}
 I have benefited a lot from numerous discussions
with my coauthors and friends A.Gerasimov,
S.Kharchev, A.Losev, Yu.Makeenko, A.Marshakov, A.Morozov, A.Orlov and
A.Zabrodin. I would also like to thank A.Marhsakov and A.Morozov for
the thoroughful reading of the text.
I am very grateful to the organizers of the XXVII Winter LIYaF School for the
hospitality. The work is
partially supported by grant 93-02-14365 of the Russian
Foundation of Fundamental Research and by NSERC (Canada).

The main
part of the lectures was written during my stay in Physics Department of the
University of British Columbia, and I highly appreciate the
hospitality and support of G.Semenoff. I would also like to thank L.Vinet for
the kind
hospitality in Centre de Recherches Mathematiques, Universite de
 Montreal, where this paper was completed.


\begin{thebibliography}{10}

\bibitem{Polyakov}
A. M. Polyakov, Phys. Lett. {\bf B103} (1981) 207, 211

\bibitem{KPZ}
V.Knizhnik, A.Polyakov, A.Zamolodchikov
 Mod.Phys.Lett. {\bf A3} (1988) 819

\bibitem{DDK}
 F.David,  Mod.Phys.Lett. {\bf A3} (1988) 1651\\
J.Distler, H.Kawai,  Nucl.Phys. {\bf B312} (1989) 509

\bibitem{GiMo}
P.Ginsparg, G.Moore, {\it
Lectures on 2D gravity and 2D string theory (TASI 1992)}, preprint
YCTP-P23-92, LA-UR-92-3479

\bibitem{dFGZ-J}
 P. Di Francesco, P. Ginsparg, and
J. Zinn-Justin, {\it $2d$ gravity and random matrices}, preprint
 LA-UR-93-1722, SPhT/93-061

\bibitem{DF}
              V.Dotsenko and V.Fateev, Nucl.Phys. {\bf B240} (1984) 312

\bibitem{Dotsenko}
Vl.Dotsenko,   Adv.Stud.in Pure Math. {\bf 16} (1988) 123

\bibitem{MarMir} A.Marshakov, A.Mironov, {\it $2d$ conformal field theories --
six years of progress}, In proceedings of
XXV Winter School of LIYaF, 1990, p.1

\bibitem{MarMor} A.Marshakov, A.Morozov, Phys.Lett. {\bf B235} (1990) 97

\bibitem{Grisaru} M.T.Grisaru, A.Lerda, S.Penati, D.Zanon, Nucl.Phys.
{\bf B342} (1990) 564

\bibitem{Zam} A.B.Zamolodchikov, JETP Lett. {\bf 43} (1986) 730

\bibitem{LianZu}
 B.Lian, G.Zuckerman,   Phys.Lett. {\bf B254} (1991) 417\\
 B.Lian, G.Zuckerman,  Phys.Lett.  {\bf B266} (1991) 21

\bibitem{BMP}
P. Bouwknegt, J. McCarthy and K. Pilch,
Comm. Math. Phys. {\bf 145} (1992) 541

\bibitem{MMMO}
 A.Marshakov, A.Mironov, A.Morozov, M.Olshanetsky,
          {\it $c=r_G$ Theories of $W_G$-Gravity: the Set of Observables as
           a {\rm Model} of Simply-Laced G},
          Preprint ITEP-M2/92, hep-th/9203043, {\it to appear in} Nucl.Phys.B.

\bibitem{Sei}
N.Seiberg, {\it Notes on quantum Liouville theory and quantum gravity},
  preprint~RU-90-29, 1990

\bibitem{BPZ} A.Belavin, A.Polyakov and A.Zamolodchikov,
                       Nucl.Phys. {\bf B241} (1984) 333

\bibitem{Wittop}
 E.Witten, Nucl.Phys. {\bf B340} (1990) 281;
     Surveys Diff.Geom. {\bf 1} (1991) 243

\bibitem{Kon}
M.Kontsevich,  Funk.Anal.\& Prilozh. {\bf 25} (1991) 50

\bibitem{DVV}
             R.Dijkgraaf, E.Verlinde and H.Verlinde, Nucl.Phys. {\bf B348}
              (1991) 565

\bibitem{DW}
R.Dijkgraaf and E.Witten, Nucl.Phys. {\bf B342} (1990) 486

\bibitem{WitN}
E.Witten, {\it The $N$ Matrix Model and Gauged WZW models},
Preprint IAS-HEP-91/26, {\it to appear in} Nucl.Phys.{\bf B}

\bibitem{Li}
                K.Li, Nucl.Phys. {\bf B354} (1991) 711

\bibitem{LGMT}
R.Dijkgraaf, E.Verlinde and H.Verlinde, Nucl.Phys.{\bf B352} (1991) 59

\bibitem{LGMTV}
E.Martinec, Phys.Lett.{\bf 217B} (1989) 431; {\it Criticalities, Catastrophe
and Compactigications}, in V.Knizhnik memorial volume, 1989; \\
C. Vafa and N.Warner, Phys.Lett.{\bf 218B} (1989) 51; \\
B.Greene, C.Vafa and N.Warner, Nucl.Phys.{\bf B324} (1989) 371; \\
W.Lerche, C.Vafa and N.Warner, Nucl.Phys.{\bf B324} (1989) 427; \\
C.Vafa, Mod.Phys.Lett. {\bf A6} (1990) 337; \\
S.Cecotti and C.Vafa, Nucl.Phys.{\bf B367} (1991) 359

\bibitem{Dij}
R.Dijkgraaf, {\it Intersection theory, integrable hierarchies
    and topological field theory}, Preprint IASSNS-HEP-91/91, hep-th 9201003

\bibitem{Los}
A.Losev, {\it Descendants constructed from matter fields and
             K.Saito higher residue pairing in topological L-G theories
      coupled to topological gravity}, Preprint TPI-MINN, {\it May}, 1992

\bibitem{LosPol}
             A.Losev and I.Polyubin, {\it On Connection between Topological
             Landau-Ginzburg Gravity
         and Integrable Systems}, Preprint ITEP/UU-ITP, Dec.1992

\bibitem{Dav}
F. David, Nucl. Phys. {\bf B257[FS14]} (1985) 45, 543\\
J. Ambj{\o}rn, B. Durhuus and J. Fr\"ohlich, Nucl. Phys. {\bf B257[FS14]}
(1985) 433\\
V. A. Kazakov, I. K. Kostov and A. A. Migdal, Phys. Lett. {\bf B157} (1985) 295

\bibitem{BK}
             E.Brezin and V.Kazakov, Phys.Lett. {\bf 236B} (1990) 144

\bibitem{DS}
             M.Douglas and S.Shenker, Nucl.Phys. {\bf B335} (1990) 635

\bibitem{GM}
             D.Gross and A.Migdal, Phys.Rev.Lett. {\bf 64} (1990) 12

 \bibitem{FKN}  M.Fukuma,  H.Kawai and  R.Nakayama, Int.J.Mod.Phys. {\bf A6}
              (1991) 1385

\bibitem{Dou} M.Douglas, Phys.Lett. {\bf 238B} (1990) 176

\bibitem{Knizh}
V.Knizhnik, UFN {\bf 159} \#3 (1989)  401
                        (Soviet Physics Uspekhi {\bf 32} (1989) 945)

\bibitem{MorUFN}
A.Morozov, UFN {\bf 162} \#8 (1992) 84
                (Soviet Physics Uspekhi {\bf 35} (1992) 671)

\bibitem{UMS}
 D.Friedan and S.Shenker, Phys.Lett. {\bf 175B} (1986) 287;\\
            N.Ishibashi, Y.Matsuo amd H.Ooguri, Mod.Phys.Lett. {\bf 2A} (1987)
            119;\\
             L.Alvarez-Gaume, C.Gomez and C.Reina, Phys.Lett. {\bf 190B}
(1987) 55; \\
             E.Witten, Comm.Math.Phys. {\bf 113} (1988) 529; \\
             A.Morozov, Phys.Lett. {\bf 196B} (1987) 325

\bibitem{BIKS}
 A.R. Its, A.G. Izergin, V.E. Korepin, N.A. Slavnov, Int.J. Mod.
Phys. {\bf B
4}, 1003 (1990).

\bibitem{Saito}
         S.Saito, Phys.Rev.Lett. {\bf 59} (1987) 1798;
         Phys.Rev. {\bf D36} (1987) 1819

\bibitem{Mor}
A.Morozov, {\it Integrability and matrix models},
Preprint ITEP-M2/93

\bibitem{Mar} A.Marshakov, {\it Integrable structures in matrix models and
physics of $2d$-gravity}, Preprint NORDITA-93/21 P, FIAN/TD-05/93

\bibitem{GMMMO} A.Gerasimov, A.Marshakov, A.Mironov, A.Morozov, A.Orlov,
             Nucl.Phys. {\bf B357} (1991) 565

\bibitem{KMMOZ} S.Kharchev, A.Marshakov, A.Mironov, A.Orlov and A.Zabrodin,
Nucl.Phys.{\bf B366} (1991) 569

\bibitem{BIZ} D.Bessis, C.Itzykson, J.-B.Zuber, {\sl Adv.Appl.Math.,}
{\bf 1} (1980) 109

\bibitem{UT} K.Ueno and K.Takasaki, Adv.Studies in Pure Math., {\bf 4} (1984)
1

\bibitem{KMMM} S.Kharchev, A.Marshakov, A.Mironov, A.Morozov,
   Nucl.Phys., {\bf B397} (1993) 379

\bibitem{MultM} C.Crnkovic, P.Ginsparg, G.Moore, {\sl Phys.Lett.,}
{\bf B237} (1990) 196\\
D.Gross, A.Migdal, {\sl Phys.Rev.Lett.,} {\bf 64} (1990) 717\\
E.Brezin, M.Douglas, V.Kazakov, S.Shenker, {\sl Phys.Lett.,}
{\bf B237} (1990) 43

\bibitem{IZ} C.Itzykson and J.-B.Zuber,
{\sl J.Math.Phys.}, {\bf 21} (1980) 411

\bibitem{KMMMP}
S.Kharchev, A.Marshakov, A.Mironov, S.Pakuliak, A.Morozov,
   {\it Conformal Matrix Models as an Alternative to Conventional Multimatrix
Models},           Preprint ITEP-M4/92, hep-th/9208044, {\it to appear in}
Nucl.Phys.B

\bibitem{MP}
                 A.Mironov and S.Pakuliak, {\it Double-Scaling Limit in the
Matrix Models of a New Type}, Preprint FIAN/TD/05-92

\bibitem{Newell} A.Newell, {\sl Solitons in mathematics and physics,}
Soc.Indust.Appl.Math. (1985)

\bibitem{MM} A.Mironov, A.Morozov, Phys.Lett. {\bf 252B} (1990) 47

\bibitem{AMJ}
             J.Ambjorn, J.Jurkiewicz and Yu.Makeenko, Phys.Lett.
                 {\bf 251B} (1990) 517

\bibitem{IM}
             H.Itoyama and Y.Matsuo, Phys.Lett. {\bf 255B} (1991) 202

\bibitem{D}
F. David, Loop equations and non-perturbative
effects in two dimensional quantum gravity,''
Mod. Phys. Lett. {\bf A5} (1990) 1019

\bibitem{MMM}
A.Marshakov, A.Mironov, A.Morozov, Phys.Lett. {\bf 265B} (1991)
99

\bibitem{FL}
V.Fateev and S.Lukyanov,  Int.J.Mod.Phys. {\bf A3} (1988) 507

\bibitem{GMMOS} A.Gerasimov, A.Marshakov, A.Morozov, M.Olshanetsky,
S.Shatashvili, {\sl Int.J.Mod.Phys.,} {\bf A5} (1990) 2495

\bibitem{Gava}
A.Marshakov, A.Mironov, A.Morozov, {\sl Mod.Phys.Lett.,}
{\bf A7} (1992) 1345

\bibitem{BIPZ} E.Brezin, C.Itzykson, G.Parisi and J.-B.Zuber,
        Comm.Math.Phys. {\bf 59} (1978) 35

\bibitem{GMMMMO} A.Gerasimov, Yu.Makeenko, A.Marshakov, A.Mironov, A.Morozov,
A.Orlov,
Mod.Phys.Lett. {\bf A6} (1991) 3079

\bibitem{BP}
C. Bachas and P.M.S. Petropoulos,
Phys. Lett. {\bf B247} (1990) 363

\bibitem{MMMM}
Yu.Makeenko, A.Marshakov, A.Mironov, A.Morozov, Nucl.Phys.
                    {\bf B356} (1991) 574

\bibitem{DJKM}
E.Date, M.Jimbo, M.Kashiwara, and T.Miwa.
\newblock In: {\it Proc.RIMS symp.Nonlinear integrable systems ---
  classical theory and quantum theory}, page~39, Kyoto, 1983

\bibitem{KS}
V.Kac and A.Schwarz, Phys.Lett. {\bf B257} (1991) 329

\bibitem{26}
M.Jimbo, T.Miwa and M.Sato, {\it Holonomic Quantum Fields},
I-V: Publ.RIMS, Kyoto Univ., {\bf 14} (1978) 223; {\bf 15} (1979) 201, 577,
871, 1531

\bibitem{Kaup}
 P.J.Hansen, D.J.Kaup,  J.Phys.Soc.Japan, {\bf 54} (1985) 4126

\bibitem{FKN2}
 M. Fukuma, H. Kawai, and R. Nakayama,
{\sl Comm.Math.Phys.,} {\bf 143} (1992) 371

\bibitem{GKMU}
A.Mironov, A.Morozov, G.Semenoff, {\it Unitary matrix integrals
in the framework of Generalized Kontsevich Model.
 1. Brezin-Gross-Witten Model,}
preprint ITEP-M6/93, FIAN/TD-16/93, UBC/S-93/93

\bibitem{Kon2}
 M.Kontsevich, Comm.Math.Phys. {\bf 147} \#1 (1992) 1

\bibitem{KMMMZ}
 S.Kharchev, A.Marshakov, A.Mironov, A.Morozov, A.Zabrodin,
Phys.Lett. {\bf 275B} (1992) 311;
                   Nucl.Phys. {\bf B380} (1992) 181

\bibitem{singer}
A.Marshakov, A.Mironov, A.Morozov,  Phys.Lett. {\bf 274B} (1992) 280

\bibitem{GN}
D.J. Gross, M. Newman, {\it Unitary and Hermitian matrices
in an external field II: the Kontsevich model and continuum Virasoro
constraints,} Princeton preprint PUPT-1282 (December, 1991)

\bibitem{Wit}
E.Witten, in Proc.of the NYC Conference, {\it June} 1991

\bibitem{MakSem}
Yu.Makeenko and G.Semenoff, Int.J.Mod.Phys.{\bf A6} (1991)
3455

\bibitem{Mik}
A.Mikhailov, {\it
Ward identities and W-constraints in Generalized Kontsevich Model,}
preprint UUITP 9/1993, FIAN/TD-03/93

\bibitem{LGM}
S.Kharchev, A.Marshakov, A.Mironov, A.Morozov, {\sl Mod.Phys.Lett.,}
{\bf A8} (1993) 1047-1061

\bibitem{ACKM}
J.Ambj{\o}rn,  L.Chekhov, C.Kristjansen, Yu.Makeenko, {\it Matrix model
calculations beyond the spherical limit}, {\sl preprint NBI-HE-92-89}
(1992)

\bibitem{GW}
D.Gross, E.Witten,
 Phys. Rev. {\bf D21}, (1980) 446

\bibitem{GKM2}
S.Kharchev, A.Marshakov, A.Mironov, A.Morozov, {\it
Generalized Kazakov-Migdal-Kontsevich Model:
group theory aspects,}
preprint UUITP-10/93, FIAN/TD-07/93, ITEP-M4/93

\bibitem{BG}
E.Brezin, D.Gross, Phys.Lett. {\bf B97} (1980) 120

\bibitem{PS}
V.Periwal, D.Shevitz,
Phys. Rev. Lett. {\bf 64}, (1990) 1326;
Nucl. Phys. {\bf B344}, (1990) 731

\bibitem{KM}
S.Kharchev, A.Mironov, Int.J.Mod.Phys. {\bf A7} (1992) 4803

\bibitem{BS}
K.A.Anagnostopoulos,
M.J.Bowick, A.S.Schwarz,
{\it Unitary Matrix Model String Equation and the Sato Grassmannian,} Preprint
SU-4238-497, 1991

\bibitem{HMNP}
T.Hollowood, L.Miramontes, A.Pasquinucci, C.Nappi,
{\it Hermitian vs. Anti-Hermitian 1-Matrix Models and their Hierarchies,}
Preprint IASSNS-HEP-91/59 and PUPT-1280, 1991

\bibitem{BMS}
M.Bowick, A.Morozov, D.Shewitz,  Nucl.Phys.{\bf B354} (1991) 496

\bibitem{DC}
C.Crnkovi\'c, M.Douglas, G.Moore,
 Nucl. Phys. {\bf B360},
(1991) 507

\end{thebibliography}
\end{document}